\documentclass[onecolumn, journal]{IEEEtran}

\usepackage{graphicx}
\usepackage{amsmath}
\usepackage{amssymb}
\usepackage{graphics}
\usepackage{latexsym}
\usepackage[dvips]{epsfig}
\usepackage[nospace]{cite}
\usepackage{subfigure}
\usepackage{setspace}
\usepackage{color}
\usepackage{tabularx}
\usepackage{url}

\usepackage{times}
\usepackage{textcomp}
\usepackage[left=0.625 in,top=0.82 in,right=0.625 in,bottom=1 in]{geometry}

\newtheorem{Definition}{Definition}
\newtheorem{Lemma}{Lemma}

\newtheorem{Proposition}[Lemma]{Proposition}
\newtheorem{Theorem}{Theorem}

\newtheorem{Remark}{Remark}

\allowdisplaybreaks[1]

\def\Pr{{\mathrm{Pr}}}
\def\E{{\mathrm E}}

\begin{document}
%
\title{\huge On the Scaling Exponent of Polar Codes for Binary-Input Energy-Harvesting Channels}
%
%
%



\author{Silas~L.~Fong,~\IEEEmembership{Member,~IEEE} and Vincent~Y.~F.~Tan,~\IEEEmembership{Senior Member,~IEEE}
\thanks{S.~L.~Fong and V.~Y.~F.~Tan were supported in part by NUS under Grant R-263-000-A98-750/133, and in part by NUS Young Investigator Award under Grant R-263-000-B37-133.}%
\thanks{S.~L.~Fong is with the Department of Electrical and Computer Engineering, National University of Singapore, Singapore 117583 (e-mail: \texttt{silas\_fong@nus.edu.sg}).}%
\thanks{V.~Y.~F.~Tan is with the Department of Electrical and Computer Engineering, National University of Singapore, Singapore 117583, and also with the Department of Mathematics, National University of Singapore, Singapore 119076 (e-mail: \texttt{vtan@nus.edu.sg}).}%
}

\maketitle

\begin{abstract}
This paper investigates the scaling exponent of polar codes for binary-input energy-harvesting (EH) channels with infinite-capacity batteries. The EH process is characterized by a sequence of i.i.d.\ random variables with finite variances. The scaling exponent~$\mu$ of polar codes for a binary-input memoryless channel (BMC)~$q_{Y|X}$ with capacity $\mathrm{C}(q_{Y|X})$ characterizes the closest gap between the capacity and non-asymptotic achievable rates in the following way: For a fixed average error probability $\varepsilon \in (0, 1)$, the closest gap between the capacity~$\mathrm{C}(q_{Y|X})$ and a non-asymptotic achievable rate~$R_n$ for a length-$n$ polar code scales as $n^{-1/\mu}$, i.e., $\min\{|\mathrm{C}(q_{Y|X})-R_n|\} = \Theta(n^{-1/\mu})$. It has been shown that the scaling exponent~$\mu$ for any binary-input memoryless symmetric channel (BMSC) with $\mathrm{C}(q_{Y|X})\in(0,1)$ lies between $3.579$ and $4.714$, where the upper bound $4.714$ was shown by an explicit construction of polar codes. Our main result shows that $4.714$ remains to be a valid upper bound on the scaling exponent for any binary-input EH channel, i.e., a BMC subject to additional EH constraints. Our result thus implies that the EH constraints do not worsen the rate of convergence to capacity if polar codes are employed.
An auxiliary contribution of this paper is that the upper bound on~$\mu$ holds for binary-input memoryless asymmetric channels.
\end{abstract}


\IEEEpeerreviewmaketitle

\flushbottom

\section{Introduction} \label{Introduction}
\subsection{Energy-Harvesting Channels}
The class of energy-harvesting (EH) channels we consider in this paper have input alphabets~$\mathcal{X}$ that are binary, output alphabets~$\mathcal{Y}$ that are finite but otherwise arbitrary, and batteries that have infinite capacities. The channel law of an EH channel is characterized by a transition matrix $q_{Y|X}$ where $X\in\mathcal{X}$ and $Y\in\mathcal{Y}$ denote the channel input and output respectively. At each discrete time $i \in \{1,2,\ldots\}$, a random amount of energy $E_i\in [0, \infty)$ arrives at the buffer and the source transmits a binary symbol $X_i\in \{0, 1\}$ such that
 \begin{equation*}
 \sum_{\ell=1}^i X_\ell \le \sum_{\ell=1}^i E_\ell   \qquad\mbox{almost surely}. 
 \end{equation*}
 This implies that the total harvested energy $\sum_{\ell=1}^i E_\ell$ must be no smaller than the ``energy" of the codeword $\sum_{\ell=1}^i X_\ell^2 = \sum_{\ell=1}^i X_\ell$  at every discrete time $i$ for transmission to take place successfully.
 We assume that $\{E_\ell\}_{\ell=1}^\infty$ are independent and identically distributed (i.i.d.) non-negative random variables, where $\E[E_1]=P$ and $\E[E_1^2]<+\infty$. The destination~$\mathrm{d}$ receives $Y_i$ from the channel output in time slot~$i$ for each $i\in\{1, 2, \ldots\}$, where $Y_i$ is correlated to $X_i$ according to the channel law, i.e., $p_{Y_i|X_i}(y_i|x_i)=q_{Y|X}(y_i|x_i)$ for all $(x_i, y_i)\in \mathcal{X}\times \mathcal{Y}$. We refer to the above EH channel as the \textit{binary-input EH channel}. It was shown by Fong, Tan and Yang~\cite{FTY15} that the capacity of the binary-input EH channel is
 \begin{equation}
\mathrm{C}(q_{Y|X};P)\triangleq \max_{p_X : \E[X]= P} I(X;Y) \qquad\mbox{bits per channel use}, \label{eqn:SaveAndTransmit}
 \end{equation}
 where $P=\E[E_1]$ is the expectation of the energy arrivals which is asymptotically the admissible peak power of the codeword $X^n$. The capacity result in~\eqref{eqn:SaveAndTransmit} was proved in~\cite{FTY15} using the save-and-transmit strategy introduced by Ozel and Ulukus \cite{Ozel:2012:AWGN} for achieving the capacity of additive white Gaussian noise (AWGN) channels.
The binary-input EH channel models practical situations where energy may not be fully available at the time of transmission and its unavailability may result in the transmitter not being able to put out the desired codeword. This model is applicable in large-scale sensor networks where each node is equipped with an EH device that collects a stochastic amount of energy. See~\cite{ulukus15} for a comprehensive review of recent advances in EH wireless communications.
\subsection{Polar Codes}
This paper investigates the scaling exponent of polar codes~\cite{MHU15} for the binary-input EH channel. The scaling exponent~$\mu$ of polar codes for a binary-input memoryless channel (BMC)~$q_{Y|X}$ with capacity
 \begin{align}
 \mathrm{C}(q_{Y|X}) = \max_{p_X} I(X;Y) \label{defI}
 \end{align}
 characterizes the closest gap between the channel capacity and a non-asymptotic achievable rate~$R_n$ in the following way: For a fixed decoding error probability $\varepsilon \in (0, 1)$, the closest gap between the capacity~$\mathrm{C}(q_{Y|X})$ and a non-asymptotic achievable rate~$R_n$ for a length-$n$ polar code scales as $n^{-1/\mu}$, i.e., $\min\{|\mathrm{C}(q_{Y|X})-R_n| \} = \Theta(n^{-1/\mu})$. It has been shown in \cite{HAU14, MHU15} that the scaling exponent of any binary-input memoryless symmetric channel (BMSC) with $\mathrm{C}(q_{Y|X})\in(0,1)$ lies between 3.579 and 4.714, where the upper bound $4.714$ was shown by an explicit construction of polar codes (see~\cite{GoldinBurshtein14} for a looser upper bound $5.702$). The scaling exponent of ternary-input memoryless symmetric channels has been studied in~\cite{GoldinBurshtein15}. It is well known that polar codes are capacity-achieving for binary-input memoryless asymmetric channels \cite{STA09, SRDR12,HondaYamamoto13,MHU14Allerton} and AWGN channels \cite{AbbeBarron11}, and it can be easily deduced from the aforementioned results that polar codes are capacity-achieving for BMCs with cost constraints. However, scaling exponents of polar codes for AWGN channels and BMCs with cost constraints have not been investigated yet. Therefore, we are motivated to study the scaling exponent of polar codes for BMCs with cost constraints, and in particular EH cost constraints.

\subsection{Main Contribution}
Our main result shows that for the binary-input EH channel which can also be viewed as a BMC subject to additional EH cost constraints, $4.714$ remains to be a valid upper bound on the scaling exponent of polar codes. Our result thus implies that the EH constraints do not worsen the rate of convergence to capacity (as quantified by the scaling exponent) if polar codes are employed.  This main result is proved by leveraging the following three existing results: scaling exponent analyses for BMSCs \cite{MHU15}, construction of polar codes designed for binary-input memoryless asymmetric channels \cite{HondaYamamoto13}, and the save-and-transmit strategy for EH channels \cite{FTY15}.
Our overarching strategy is to design the energy-saving phase to be sufficiently short so as not to affect the scaling exponent, yet long enough so that the error probability of the resultant code is not severely degraded relative to the case without EH constraints. An auxiliary contribution of this paper is that $4.714$ is also an upper bound on the scaling exponent of polar codes for binary-input memoryless asymmetric channels.

The main difficulty in this work is extracting and modifying the key elements in the three aforementioned works \cite{MHU15,HondaYamamoto13,FTY15} which are themselves presented under different settings. We have to perform several non-trivial modifications so that the techniques and results in \cite{MHU15,HondaYamamoto13,FTY15} can be applied to our problem. More specifically, the three different settings can be briefly described as follows: (i) The scaling exponent analyses in~\cite{MHU15} are performed for \emph{symmetric} channels rather than asymmetric channels; (ii) The polar codes designed for asymmetric channels in~\cite{HondaYamamoto13} are fixed-rate codes under the \emph{error exponent} regime rather than fixed-error codes under the scaling exponent regime; (iii) The save-and-transmit codes used in~\cite{FTY15} are \emph{random} codes with i.i.d.\ codewords (where each codeword consists of i.i.d.\ symbols) rather than structured codes.
\subsection{Paper Outline}
This paper is organized as follows. The notation used in this paper is described in the next subsection. Section~\ref{sectionDefinition} states the formulation of the binary-input EH channel, save-and-transmit polar codes and scaling exponents and presents our main theorem. Section~\ref{sectionMainResultProof} proves our main theorem, which states that 4.714 is an upper bound on the scaling exponent of save-and-transmit polar codes for the binary-input EH channel. Concluding remarks are provided in Section~\ref{sec:conclusion}.

 \subsection{Notation} \label{sectionNotation}
We let $\boldsymbol{1}\{\mathcal{E}\}$ be the indicator  function of the set  $\mathcal{E}$.
An arbitrary (discrete or continuous) random variable is denoted by an upper-case letter (e.g., $X$), and the realization and the alphabet of the random variable are denoted by the corresponding lower-case letter (e.g., $x$) and calligraphic letter (e.g., $\mathcal{X}$) respectively.
We use $X^n$ to denote the random tuple $(X_1, X_2, \ldots, X_n)$ where each $X_i$ has the same alphabet $\mathcal{X}$. We will take all logarithms to base $2$ throughout this paper unless specified otherwise. The logarithmic functions to base~$2$ and base~$e$ are denoted by $\log$ and $\ln$ respectively.  The set of natural numbers, real numbers and non-negative real numbers are denoted by $\mathbb{N}$, $\mathbb{R}$ and $\mathbb{R}_+$ respectively.

The following notations are used for any arbitrary random variables~$X$ and~$Y$ and any real-valued function $g$ with domain $\mathcal{X}$. We let $p_{Y|X}$ and $p_{X,Y}=p_Xp_{Y|X}$ denote the conditional probability distribution of $Y$ given $X$ and the probability distribution of $(X,Y)$ respectively.
We let $p_{X,Y}(x,y)$ and $p_{Y|X}(y|x)$ be the evaluations of $p_{X,Y}$ and $p_{Y|X}$ respectively at $(X,Y)=(x,y)$. 
To make the dependence on the distribution explicit, we let $\Pr_{p_X}\{ g(X)\in\mathcal{A}\}$ denote $\int_{x\in \mathcal{X}} p_X(x)\mathbf{1}\{g(x)\in\mathcal{A}\}\, \mathrm{d}x$ for any set $\mathcal{A}\subseteq \mathbb{R}$.
The expectation of~$g(X)$ is denoted as
$\E_{p_X}[g(X)]$. For any $(X,Y, Z)$ distributed according to some $p_{X,Y, Z}$, the entropy of $X$ and the conditional mutual information between $X$ and $Y$ given~$Z$ are denoted by $H_{p_X}(X)$ and $I_{p_{X,Y,Z}}(X;Y|Z)$ respectively. For simplicity, we sometimes omit the subscript of a notation if it causes no confusion. The total variation distance between $p_X$ and $q_X$ is denoted by \[
\|p_X-q_X\|\triangleq \frac{1}{2}\sum_{x\in \mathcal{X}}|p_X(x)-q_X(x)|.
 \]

\section{Problem Formulation, Preliminaries and \\ Main Result}
\label{sectionDefinition}
\subsection{Binary-Input EH Channel}
We follow the formulation of EH channels in~\cite{FTY15}. The binary-input EH channel consists of one source and one destination, denoted by $\mathrm{s}$ and $\mathrm{d}$ respectively. Node~$\mathrm{s}$ transmits information to node~$\mathrm{d}$ in $n$ time slots as follows. Node~$\mathrm{s}$ chooses message
$
W
$
 and sends $W$ to node~$\mathrm{d}$, where $W$ is uniformly distributed over $\{1, 2, \ldots, M\}$ and $M$ denotes the message size. The energy-harvesting process is characterized by $E_1, E_2, \ldots, E_n$, which are i.i.d.\ real-valued random variables that satisfy
$
\Pr\{E_1< 0\}=0$,
$
\E[E_1]=P$
and
$
\E[E_1^2]<\infty$. Then for each $i\in \{1, 2, \ldots, n\}$, node~$\mathrm{s}$ transmits $X_i\in \{0,1\}$ based on~$(W, E^i)$ and node~$\mathrm{d}$ receives $Y_i\in \mathcal{Y}$ in time slot~$i$ where $\mathcal{Y}$ is an arbitrary finite alphabet.
We assume the following for each $i\in\{1, 2, \ldots, n\}$:
\begin{enumerate}
\item[(i)] $E_i$ and $(W, E^{i-1}, X^{i-1}, Y^{i-1})$ are independent, i.e.,
\begin{align}
p_{W, E^i, X^{i-1}, Y^{i-1}} = p_{E_i}p_{W, E^{i-1}, X^{i-1}, Y^{i-1}}. \label{assumption(i)}
\end{align}
    \item[(ii)] Every codeword $X^n$ transmitted by~$\mathrm{s}$ satisfies
\begin{equation}
\Pr\left\{\sum_{\ell=1}^i X_\ell \le \sum_{\ell=1}^i E_\ell\right\}=1 \label{eqn:eh}
\end{equation}
for each $i\in\{1, 2, \ldots, n\}$.
  \end{enumerate}
   After~$n$ time slots, node~$\mathrm{d}$ declares~$\hat W$ to be the transmitted~$W$ based on $Y^n$. Formally, we define a code as follows:
\begin{Definition} \label{defCode}
An {\em $(n, M)$-code} consists of the following:
\begin{enumerate}
\item A message set
$
\mathcal{W}\triangleq \{1, 2, \ldots, M\}
$
 at node~$\mathrm{s}$. Message $W$ is uniform on $\mathcal{W}$.

\item A sequence of encoding functions
$
f_i : \mathcal{W}\times \mathbb{R}_+^i\rightarrow \{0,1\}
$
 for each $i\in\{1, 2, \ldots, n\}$, where $f_i$ is the encoding function for node~$\mathrm{s}$ at time slot~$i$ for encoding $X_i$ such that~$
X_i=f_i (W, E^i).
$
\item A decoding function
$
\varphi :
\mathbb{R}^{n} \rightarrow \mathcal{W},
$
for decoding $W$ at node~$\mathrm{d}$ by producing
$
 \hat W = \varphi(Y^{n})$.
\end{enumerate}
If the sequence of encoding functions $f_i$ satisfies the EH constraints~\eqref{eqn:eh}, the code is also called an {\em $(n, M)$-EH code}.
\end{Definition}
\medskip

By Definition~\ref{defCode}, the only potential difference between an $(n, M)$-EH code and an $(n, M)$ code is whether the EH constraints~\eqref{eqn:eh} are satisfied. If an $(n, M)$-code does not satisfy the EH constraints~\eqref{eqn:eh} during the encoding process (i.e., $X^n$ is a function of~$W$ alone), then the $(n, M)$-code can be viewed as an $(n, M)$-code for the usual discrete memoryless channel (DMC) without any cost constraint~\cite[Sec.~3.1]{elgamal}.
\medskip
\begin{Definition}\label{defChannel}
The {\em binary-input EH channel} is characterized by a binary input alphabet~$\mathcal{X}\triangleq \{0, 1\}$, a finite output alphabet~$\mathcal{Y}$ and a transition matrix $q_{Y|X}$ such that the following holds for any $(n, M)$-code: For each $i\in\{1, 2, \ldots, n\}$,
\begin{align*}
p_{W, E^i, X^i, Y^i}
 = p_{W, E^i, X^i, Y^{i-1}}p_{Y_i|X_i} 
\end{align*}
where
\begin{equation}
p_{Y_i|X_i}(y_i|x_i) = q_{Y|X}(y_i|x_i) \label{defChannelInDefinition*}
\end{equation}
for all $x_i\in \mathcal{X}$ and $y_i\in \mathcal{Y}$.
Since $p_{Y_i|X_i}$ does not depend on~$i$ by \eqref{defChannelInDefinition*}, the channel is stationary.
\end{Definition}
\medskip
\begin{Definition}
The binary-input channel $q_{Y|X}$ is said to be \textit{symmetric} if there exists a permutation $\pi$ of the output alphabet $\mathcal{Y}$ such that (i) $\pi^{-1}=\pi$ and (ii) $q_{Y|X}(y|1)=q_{Y|X}(\pi(y)|0)$ for all $y\in \mathcal{Y}$. Otherwise, the channel is said to be \textit{asymmetric}.
\end{Definition}
\medskip

 For any $(n, M)$-code defined on the binary-input EH channel, let $p_{W,E^n, X^n, Y^n, \hat W}$ be the joint distribution induced by the code. We can factorize $p_{W,E^n, X^n, Y^n, \hat W}$ as
\begin{align}
 p_{W,E^n, X^n, Y^n, \hat W}
=p_W \left(\prod_{i=1}^n p_{E_i} p_{X_i|W, E^i} p_{Y_i|X_i}\right)p_{\hat W |Y^n}, \label{memorylessStatement}
\end{align}
which follows from the i.i.d.\ assumption of the EH process $E^n$ in~\eqref{assumption(i)}, the fact by Definition~\ref{defCode} that $X_i$ is a function of $(W, E^i)$ and the memorylessness of the channel $q_{Y|X}$ described in Definition~\ref{defChannel}.
\smallskip
\begin{Definition} \label{defErrorProbability}
For an $(n, M)$-code defined on the binary-input EH channel, we can calculate according to \eqref{memorylessStatement} the \textit{average probability of decoding error} defined as $\Pr\big\{\hat W \ne W\big\}$.
We call an $(n, M)$-code and an $(n, M)$-EH code with average probability of decoding error no larger than~$\varepsilon$ an {\em $(n, M, \varepsilon)$-code} and an {\em $(n, M, \varepsilon)$-EH code} respectively.
\end{Definition}
\smallskip
\begin{Definition} \label{defAchievableRate}
Let $\varepsilon\in (0,1)$ be a real number. A rate $R$ is said to be \textit{$\varepsilon$-achievable} for the EH channel if there exists a sequence of $(n, M_n, \varepsilon)$-EH codes such that
\begin{equation*}
\liminf_{n\rightarrow \infty}\frac{1}{n}\log M_n \ge R.
\end{equation*}
\end{Definition}

\begin{Definition}\label{defCapacity}
Let $\varepsilon\in (0,1)$ be a real number. The {\em $\varepsilon$-capacity} of the binary-input EH channel, denoted by $C_\varepsilon$, is defined to be
$
C_\varepsilon \triangleq \sup\{R: R\text{ is $\varepsilon$-achievable for the EH channel}\}$. The \emph{capacity} of the binary-input EH channel is $C\triangleq \inf_{\varepsilon>0}C_\varepsilon $.
\end{Definition}
\medskip

Define the capacity-cost function
\begin{equation}
\mathrm{C}(q_{Y|X};P) \triangleq
\max\limits_{p_X: \E_{p_X}[X]=P}I_{p_X q_{Y|X}}(X;Y) \label{defCDMC}.
\end{equation}
It was shown in \cite[Sec.~IV]{FTY15} that
 \begin{equation*}
 C_\varepsilon = C = \mathrm{C}(q_{Y|X};P)
 \end{equation*}
 for all $\varepsilon\in (0,1)$.
 The following proposition is a direct consequence of\cite[Lemma~4]{FTY15}, which will be useful for calculating the length of energy-saving phase for the save-and-transmit strategy.
\medskip
\begin{Proposition} \label{propositionCharacteristicFunctionDMC}
Let $m$ and $n$ be two natural numbers. Suppose $\{X_i\}_{i=1}^n$ and $\{E_i\}_{i=1}^{m+n}$ are two sequences of i.i.d.\ random variables such that $X_1\in \{0,1\}$, $X^n$ and $E^{m+n}$ are independent,
\begin{equation*}
\Pr_{p_{E_1}}\{E_1 < 0\}=0, 
\end{equation*}
and
\begin{equation*}
\E_{p_{E_1}}[E_1]=\E_{p_{X_1}}[X_1]=P. 
\end{equation*}
In addition, suppose $\E_{p_{E_1}}[E_1^2]<\infty$ and define
\begin{equation}
a\triangleq \max\left\{\E_{p_{E_1}}[E_1^2], e \right\}. \label{propositionCharFuncAssump2DMC}
\end{equation}
If $n\ge 3$ is sufficiently large such that
\begin{equation*}
\frac{n}{\ln n}\ge \frac{a}{P^2}\, , 
\end{equation*}
then we have
\begin{align}
\Pr_{p_{X^n}p_{E^{m+n}}}\left\{\bigcup_{i=1}^n \left\{\sum_{\ell=1}^i X_\ell \ge \sum_{\ell=1}^{m+i} E_\ell\right\}\right\} \le  \left(\frac{e^{0.4} }{\ln n}\right) e^{2\ln n -\frac{mP}{2}\sqrt{\frac{\ln n}{a n}}}. \label{stPropositionCharFunction}
\end{align}
\end{Proposition}
\begin{IEEEproof}
Proposition~\ref{propositionCharacteristicFunctionDMC} follows from \cite[Lemma~4]{FTY15} by letting $c(x)=x$ for each $x\in\{0,1\}$.
\end{IEEEproof}
\begin{Remark}
Proposition~\ref{propositionCharacteristicFunctionDMC} implies that if the source harvests energy for~$m$ channel uses before transmitting a random codeword $X^n$ consisting of i.i.d.\ symbols, then the probability that $X^n$ violates the EH constraint (cf.\ \eqref{eqn:eh}) is bounded above as~\eqref{stPropositionCharFunction}.
\end{Remark}

\subsection{Polarization for Binary Memoryless Asymmetric Channels}
We follow the formulation of polar coding in~\cite{HondaYamamoto13}. For any tuple of discrete random variables $(U,X,Y)$ distributed on $\mathcal{U}\times \mathcal{X}\times \mathcal{Y}$ according to $p_{U,X,Y}$ where $\mathcal{U}=\{0,1\}$, the corresponding Bhattacharyya parameter is defined to be
\begin{align}
Z_{p_{U, X, Y}}(U|Y) & \triangleq  2\sum_{y\in \mathcal{Y}}p_{Y}(y)\sqrt{p_{U|Y}(0|y)p_{U|Y}(1|y)}\notag\\
& = 2\sum_{y\in \mathcal{Y}}\sqrt{p_{U,Y}(0, y)p_{U,Y}(1,y)}, \label{defBhattacharyya}
  \end{align}
  where $p_Y$, $p_{U|Y}$ and $p_{U,Y}$ are marginal distributions of $p_{U, X, Y}$. It is well known that \cite[Proposition~2]{Arikan:10ISIT}
  \begin{equation}
(Z_{p_{U, X, Y}}(U|Y))^2 \le H_{p_{U,X,Y}}(U|Y). \label{bhattacharyyaEntropy}
  \end{equation}
  Let $p_X$ be the probability distribution of a Bernoulli random variable~$X$, and let $p_{X^n}$ be the distribution of~$n$ independent copies of~$X\sim p_X$, i.e., $p_{X^n}(x^n) = \prod_{i=1}^n p_{X}(x_i)$ for all $x^n\in \mathcal{X}^n$. For $n=2^k$ for each $k\in \mathbb{N}$, the polarization mapping of polar codes is given by
  \begin{equation}
  G_n \triangleq \bigg[\begin{matrix} 1& 0  \\  1 & 1 \end{matrix}\bigg]^{\otimes k} = G_n^{-1} \label{defGn}
  \end{equation}
  where $\otimes$ denotes the Kronecker power. Define $p_{U^n|X^n}$ such that
  \begin{equation}
[U_1\ U_2\ \ldots \ U_n]  = [X_1\ X_2\ \ldots \ X_n] G_n, \label{defpUgivenX}
  \end{equation}
define
  \begin{equation*}
  p_{Y_i|X_i}(y_i|x_i) \triangleq q_{Y|X}(y_i|x_i)
  \end{equation*}
  for each $i\in\{1, 2, \ldots, n\}$ and each $(x_i, y_i)\in \mathcal{X}\times \mathcal{Y}$ where $q_{Y|X}$ is the channel transition matrix (cf.\ \eqref{defChannel}),
  and define
  \begin{equation}
  p_{U^n, X^n, Y^n}\triangleq p_{X^n} p_{U^n|X^n}\prod_{i=1}^n p_{Y_i|X_i}. \label{defP}
  \end{equation}
\medskip

The following lemma is useful for establishing our scaling exponent upper bound for the binary-input EH channel. The proof combines key ideas in~\cite{MHU15} and~\cite{HondaYamamoto13}, and is relegated to Appendix~\ref{appendixA}.
\medskip
\begin{Lemma}\label{lemmaPolar}
Let $\mu=4.714$. For any binary-input channel $q_{Y|X}$ and any $p_X$, define $p_{U^n, X^n, Y^n}$ as in~\eqref{defP} for each $n\in\mathbb{N}$. Then, there exist two positive numbers~$t_1$ and~$t_2$ which do not depend on~$n$ such that for any $k\in\mathbb{N}$ and $n\triangleq 2^k$, we have\footnote{This lemma remains to hold if the quantities $\frac{1}{n^4}$ are replaced by $\frac{1}{n^\nu}$ for any $\nu>0$. The main result of this paper continues to hold if the quantities $\frac{1}{n^4}$ in this lemma are replaced by $\frac{1}{n^\nu}$ for any $\nu > 2$.}
 \begin{align}
 \frac{1}{n}\left|\left\{ i\in\{1, 2, \ldots, n\}\left|\parbox[c]{1.9 in}{$ Z_{p_{U^n, X^n, Y^n}}(U_i|U^{i-1}, Y^n) \le \frac{1}{n^4}, \vspace{0.04 in} \\
 Z_{p_{U^n, X^n, Y^n}}(U_i|U^{i-1}) \ge 1-\frac{1}{n^4}$}  \right.  \right\}\right| \ge I_{p_X q_{Y|X}}(X; Y) - \frac{t_1}{n^{1/\mu}}. \label{st1InLemmaPolar}
 \end{align}
and
  \begin{align*}
 \frac{1}{n}\!\left|\left\{ i\in\{1, 2, \ldots, n\}\!\left| \parbox[c]{2.08 in}{$ Z_{p_{U^n, X^n, Y^n}}(U_i|U^{i-1}, Y^n) \ge 1-  \frac{1}{n^4}, \\Z_{p_{U^n, X^n, Y^n}}(U_i|U^{i-1}) \le \frac{1}{n^4}$} \!\!\right.  \right\}\right| \ge 1- I_{p_X q_{Y|X}}(X; Y) - \frac{t_2}{n^{1/\mu}}. 
 \end{align*}
\end{Lemma}

\begin{Remark}
The bound in~\eqref{st1InLemmaPolar} in Lemma~\ref{lemmaPolar} tells us that the fraction of good synthesized channels in terms of their Bhattachryya parameters is close to the mutual information $I(X;Y)$. Furthermore the notions of ``good" and ``close to $I(X;Y)$" are quantified precisely as functions of the blocklength. These quantifications of the rates of convergence allow us to establish a meaningful bound on the scaling exponent.
\end{Remark}

\subsection{Definitions of Polar Codes}
The following definition of polar codes is motivated by Lemma~\ref{lemmaPolar} and the construction of polar codes in~\cite[Sec.~III-A]{HondaYamamoto13}.
\medskip
\begin{Definition} \label{defPolarCode}
Fix a $k\in\mathbb{N}$, and let $n=2^k$. For any binary-input channel $q_{Y|X}$ and any $p_X$, define $p_{U^n, X^n, Y^n}$ as in~\eqref{defP}. Let $\mathcal{I}\subseteq \{1, 2, \ldots, n\}$ be a set to be specified shortly and fix a collection of functions $\lambda_i: \{0, 1\}^{i-1}\rightarrow \{0,1\}$ for each $i\in \mathcal{I}^c$. An $(n, p_X,  \mathcal{I}, \lambda_{\mathcal{I}^c})$-polar code with
$
\lambda_{\mathcal{I}^c}\triangleq (\lambda_i |\, i\in \{1, 2, \ldots, n\}\setminus \mathcal{I})
$
 consists of the following:
\begin{enumerate}
\item An index set for information bits
\begin{equation}
\mathcal{I}\!\triangleq\! \left\{ i\in \{1, 2, \ldots, n\}\left|\parbox[c]{1.9 in}{$ Z_{p_{U^n, X^n, Y^n}}(U_i|U^{i-1}, Y^n) \!\le \! \frac{1}{n^4}, \\
 Z_{p_{U^n, X^n, Y^n}}(U_i|U^{i-1}) \ge 1-\frac{1}{n^4}$}  \!\!\!\right.  \right\}. \label{defInformationBitSet}
\end{equation}
The set
 \begin{equation}
 \mathcal{I}^c \triangleq \{1, 2, \ldots, n\}\setminus \mathcal{I} \label{defFrozenBitSet}
 \end{equation}
 is referred to as the index set for frozen bits.

 \item A message set $\mathcal{W}\triangleq \{1, 2, \ldots, 2^{|\mathcal{I}|}\}$, where~$W$ is uniform on~$\mathcal{W}$.

 \item An encoding bijection $f: \mathcal{W} \rightarrow \mathcal{U}_{\mathcal{I}}$ for information bits denoted by $U_{\mathcal{I}}$ such that
   \begin{equation*}
  U_{\mathcal{I}} = f(W),
   \end{equation*}
   where $\mathcal{U}_{\mathcal{I}}$ and $  U_{\mathcal{I}}$ are defined as $\mathcal{U}_{\mathcal{I}} \triangleq \prod_{i\in\mathcal{I}}\mathcal{U}_i$ and $ U_{\mathcal{I}} \triangleq (U_i|i\in\mathcal{I})$ respectively.
   Since the message is uniform on~$\mathcal{W}$, $f(W)$ is a sequence of uniform i.i.d.\ bits such that
   \begin{equation}
 \Pr\{U_{\mathcal{I}} = u_{\mathcal{I}}\}=\frac{1}{2^{|\mathcal{I}|}} \label{defUniformBits}
   \end{equation}
   for all $u_{\mathcal{I}}\in\{0,1\}^{|\mathcal{I}|}$, where the bits are transmitted through the polarized channels indexed by~$\mathcal{I}$.

\item For each $i\in \mathcal{I}^c$, an encoding function $\lambda_i:\{0, 1\}^{i-1}\rightarrow \{0,1\}$ for frozen bit~$U_i$ such that
\begin{equation}
U_i = \lambda_i(U^{i-1}). \label{defLambdaI}
\end{equation}
After $U^n$ has been determined, node~$\mathrm{s}$ transmits $X^n$ where
\begin{equation}
[X_1\ X_2\ \ldots \ X_n] \triangleq [U_1\ U_2\ \ldots \ U_n]G_n^{-1}.  \label{defpUgivenXinPolar}
\end{equation}
If the encoding functions $\lambda_{\mathcal{I}^c}$ for the frozen bits are stochastic (which we allow), then they will also be denoted by $\Lambda_{\mathcal{I}^c}$ for clarity.

\item A sequence of successive cancellation decoding functions $\varphi_i: \{0, 1\}^{i-1}\times \mathcal{Y}^n \rightarrow \{0,1\}$ for each $i\in\{1, 2, \ldots, n\}$ such that the recursively generated $\hat U_1, \hat U_2, \ldots, \hat U_n$ are produced as follows for each $i=1, 2, \ldots, n$:
    \begin{equation*}
    \hat U_i \triangleq \varphi_i(\hat U^{i-1}, Y^n)
    \end{equation*}
    where
    \begin{align}
  \hat u_i &\triangleq  \varphi_i(\hat u^{i-1}, y^n) \notag\\*
  &=
    \begin{cases}
  0  & \text{if $i\in\mathcal{I}$ and $p_{U_i|U^{i-1}, Y^n}(0|\hat u^{i-1} \!, y^n) \ge p_{U_i|U^{i-1}, Y^n}(1|\hat u^{i-1}, y^n)$,}\vspace{0.04 in}\\
   1 & \text{if $i\in\mathcal{I}$ and $p_{U_i|U^{i-1}, Y^n}(0|\hat u^{i-1} \!, y^n)  < p_{U_i|U^{i-1}, Y^n}(1|\hat u^{i-1}, y^n)$,}\vspace{0.04 in} \\
   \lambda_i(\hat u^{i-1}) & \text{if $i\in\mathcal{I}^c$.}
    \end{cases}
    \label{defSCdecoder}
    \end{align}
    After obtaining $\hat U^n$, the estimate of $U^n$, node~$\mathrm{d}$ declares that
    \begin{equation*}
    \hat W \triangleq f^{-1}(\hat U^n)
    \end{equation*}
    is the transmitted message.
\end{enumerate}
\end{Definition}
\begin{Remark}
By inspecting Definition~\ref{defCode} and Definition~\ref{defPolarCode}, we see that every $(n, p_X,  \mathcal{I}, \lambda_{\mathcal{I}^c})$-polar code is also an~$(n, 2^{|\mathcal{I}|})$-code.
\end{Remark}
\begin{Remark} \label{remark2}
For any $(n, p_X,  \mathcal{I}, \lambda_{\mathcal{I}^c})$-polar code as defined in Definition~\ref{defPolarCode}, although the Bhattacharyya parameters \linebreak $Z_{p_{U^n, X^n, Y^n}}(U_i|U^{i-1}, Y^n)$ and
 $Z_{p_{U^n, X^n, Y^n}}(U_i|U^{i-1})$ are calculated according to $p_{U^n, X^n, Y^n}$ where $p_{X^n}(x^n)=\prod_{i=1}^n p_X(x_i)$ and $p_{U^n|X^n}$ characterizes the polarization mapping according to~\eqref{defpUgivenX}, the distribution induced by the polar code is not equal to $p_{U^n, X^n, Y^n}$. Indeed, the distribution induced by the polar code depends on the uniform i.i.d.\ information bits $U_\mathcal{I}$, the encoding functions~$\lambda_{\mathcal{I}^c}$ of the frozen bits $U_{\mathcal{I}^c}$, the polarization map $G_n$ defined in~\eqref{defGn} and the channel law $q_{Y|X}$.
\end{Remark}
\medskip

\begin{Definition}\label{defErrorProbabilityPolar}
For an $(n, p_X,  \mathcal{I}, \lambda_{\mathcal{I}^c})$-polar code, the probability of decoding error is defined as
\begin{equation*}
\Pr\{\hat W \ne W\} = \Pr\{\hat U_{\mathcal{I}} \ne U_{\mathcal{I}}\}
\end{equation*}
where the error is averaged over the random message as well as the potential randomness of $\lambda_{\mathcal{I}^c}$ (which could be stochastic).
The code is also called an $(n, p_X,  \mathcal{I}, \lambda_{\mathcal{I}^c}, \varepsilon)$-polar code if the probability of decoding error is no larger than $\varepsilon$.
\end{Definition}
\medskip
\begin{Remark}
For an $(n, p_X,  \mathcal{I}, \lambda_{\mathcal{I}^c})$-polar code, although the Bhattacharyya parameters are evaluated according to $p_{U^n, X^n, Y^n}$ as defined in~\eqref{defP}, the probability terms in Definition~\ref{defErrorProbabilityPolar} are evaluated according to the distribution induced by the code, which is not $p_{U^n, X^n, Y^n}$ as explained in Remark~\ref{remark2}.
\end{Remark}

\subsection{Definitions for the EH Transmission Strategy}
In this paper, we investigate the save-and-transmit strategy in~\cite{FTY15} for polar codes under the EH constraints~\eqref{eqn:eh}, which is formally defined as follows.
\medskip
\begin{Definition} \label{defPolarCodeEH}
Let $m$ and $n$ be two non-negative integers such that $n=2^k$ for some $k\in\mathbb{N}$. A \emph{save-and-transmit} $(m, (n, p_X,  \mathcal{I}, \lambda_{\mathcal{I}^c}))$-EH polar code consists of the following:
\begin{enumerate}
%
\item An energy-harvesting period of~$m$ time slots in which node~$\mathrm{s}$ always transmits~$0$ and
 a transmission period of~$n$ time slots in which node~$\mathrm{s}$ tries to transmit information.

\item A message set $\mathcal{W}\triangleq \{1, 2, \ldots, 2^{|\mathcal{I}|}\}$, where $\mathcal{I} \subseteq \{1, 2, \ldots, n\}$ and~$W$ is uniform on~$\mathcal{W}$.

\item An $(n, p_X,  \mathcal{I}, \lambda_{\mathcal{I}^c})$-polar code (as described in Definition~\ref{defPolarCode}) with an encoding bijection $\tilde f: \mathcal{W} \rightarrow \mathcal{U}_{\mathcal{I}}$ for information bits denoted by~$U_{\mathcal{I}}$, an encoding function $\lambda_i:\{0, 1\}^{i-1}\rightarrow \{0,1\}$ for frozen bit $U_i$ for each $i\in \{1, 2, \ldots, n\}\setminus\mathcal{I}$ and a sequence of successive cancellation decoding functions $\tilde{\varphi}_i: \{0, 1\}^{i-1}\times \mathcal{Y}^{n} \rightarrow \{0,1\}$ for each $i\in\{1, 2, \ldots, n\}$. Let
    \begin{align}
    [\tilde X_1\ \tilde X_2\ \ldots \tilde X_{n}] \triangleq [ U_1\  U_2\ \ldots  U_{n}] G_{n}^{-1} \label{polarizationInvMap}
    \end{align}
   be the $n$ transmitted symbols induced by the $(n, p_X,  \mathcal{I}, \lambda_{\mathcal{I}^c})$-polar code, where the distribution of $U^{n}$ is fully determined by the uniformity of message~$W$, the bijection $\tilde f$ and the sequence of $\lambda_i$.

\item A sequence of encoding functions $f_i: \mathcal{W}\times \mathbb{R}_+^i \rightarrow \mathcal{X}$ that intends to transmit codewords of the $(n, p_X, \mathcal{I}, \lambda_{\mathcal{I}^c})$-polar code during the transmission period subject to the EH constraints~\eqref{eqn:eh}, where the symbol transmitted in time slot~$i$ is
    \begin{align}
    f_i(W, E^i) \triangleq\begin{cases} 0 & \text{if $1\le i\le m$,} \vspace{0.04 in}\\  \tilde X_{i-m} & \text{if $m+1 \le i \le m+n$ and  ${ }\: \tilde X_{i-m} \le \sum_{\ell=1}^i E_\ell - \sum_{\ell=1}^{i-1} f_\ell(W,E^{\ell}) $,}\vspace{0.04 in}\\ 0 & \text{if $m+1 \le i \le m+n$ and  ${ }\: \tilde X_{i-m} > \sum_{\ell=1}^i E_\ell - \sum_{\ell=1}^{i-1} f_\ell(W,E^{\ell}) $.}\end{cases} \label{defFiPolar}
    \end{align}
    By~\eqref{defFiPolar}, the EH constraint
    \begin{equation}
  \sum_{\ell=1}^{i} f_\ell(W,E^{\ell})\le \sum_{\ell=1}^i E_\ell \label{EHconstraintPolar}
    \end{equation}
    is satisfied for each $i\in\{1, 2, \ldots, m+n\}$.
    Let $Y^{m+n}$ be the symbols received by node~$\mathrm{d}$ during the~$m+n$ time slots, and let
    \begin{equation*}
    \tilde Y^{n} \triangleq (Y_{m+1}, Y_{m+2}, \ldots, Y_{m+n})
    \end{equation*}
    be the symbols received by node~$\mathrm{d}$ during the transmission period.
    \item A sequence of successive cancellation decoding functions $\varphi_i: \{0, 1\}^{i-1}\times \mathcal{Y}^{n} \rightarrow \{0,1\}$ for each $i\in\{1, 2, \ldots, n\}$ such that the recursively generated $\hat U_1, \hat U_2, \ldots, \hat U_{n}$ are produced as follows for each $i=1, 2, \ldots, n$:
    \begin{equation*}
    \hat U_i \triangleq \varphi_i(\hat U^{i-1}, \tilde Y^{n})
    \end{equation*}
    where
    \begin{align}
  \varphi_i(\hat u^{i-1}, \tilde y^{n}) \triangleq \tilde{\varphi}_i(\hat u^{i-1}, \tilde y^{n}). \label{defSCdecoderEH}
    \end{align}
    After obtaining $\hat U^{n}$, the estimates of $U^{n}$, node~$\mathrm{d}$ declares that
    \begin{equation*}
    \hat W \triangleq \tilde f^{-1}(\hat U^{n})
    \end{equation*}
    is the transmitted message.
\end{enumerate}
\end{Definition}
\medskip

The $(n, p_X,  \mathcal{I}, \lambda_{\mathcal{I}^c})$-polar code described in Definition~\ref{defPolarCodeEH} is called the \emph{effective code} of the save-and-transmit $(m, (n, p_X,  \mathcal{I}, \lambda_{\mathcal{I}^c}))$-EH polar code. By Definition~\ref{defPolarCodeEH}, the effective code of the $(m, (n, p_X,  \mathcal{I}, \lambda_{\mathcal{I}^c}))$-EH polar code fully determines the encoding and decoding functions of the save-and-transmit EH polar code, where the latter polar code ensures that the EH constraints to be satisfied. In addition, if the overall probability of decoding error is no larger than~$\varepsilon$, i.e.,
    \begin{align*}
    \Pr\{\hat W \ne W\}=    \Pr\{\hat U^{n} \ne U^{n}\} \le \varepsilon,
    \end{align*}
    where the error is averaged over the random message as well as the potential randomness of $\lambda_{\mathcal{I}^c}$, then the code is also called a \emph{save-and-transmit $(m, (n, p_X,  \mathcal{I}, \lambda_{\mathcal{I}^c}), \varepsilon)$-EH polar code.}
\medskip
\begin{Remark}
By inspecting Definitions~\ref{defCode}, \ref{defErrorProbability}, \ref{defPolarCode} and~\ref{defPolarCodeEH}, we see that any save-and-transmit $(m, (n, p_X,  \mathcal{I}, \lambda_{\mathcal{I}^c}), \varepsilon)$-EH polar code is also an~$(m+n, 2^{|\mathcal{I}|}, \varepsilon)$-EH code.
\end{Remark}

\subsection{Scaling Exponent}
\begin{Definition} \label{defScalingExp}
Fix an~$\varepsilon\in(0,1)$ and a BMC $q_{Y|X}$ with capacity $\mathrm{C}(q_{Y|X})\in (0,1)$ (defined in~\eqref{defI}).
The \emph{scaling exponent of polar codes for the BMC} is defined as
\begin{align*}
 \mu_\varepsilon^{\text{\tiny PC-BMC}} \triangleq \liminf_{n\rightarrow \infty} \inf \left\{\left.\frac{-\log n}{\log \left|\mathrm{C}(q_{Y|X})- \frac{|\mathcal{I}|}{n}\right|}\right| \parbox[c]{1.18 in}{There exists an \\$(n, p_X,  \mathcal{I}, \lambda_{\mathcal{I}^c}, \varepsilon)$-polar code on $q_{Y|X}$}\right\}.
\end{align*}
\end{Definition}

Definition~\ref{defScalingExp} formalizes the notion that we are seeking the smallest $\mu\ge 0$ such that $|\mathrm{C}(q_{Y|X})-R_n|=O(n^{-1/\mu})$ holds.
It has been shown in \cite[Sec.~IV-C]{HAU14} and \cite[Th.~2]{MHU15} that
\begin{equation}
3.579 \le \mu_\varepsilon^{\text{\tiny PC-BMC}}\le 4.714  \qquad \forall \varepsilon\in(0,1)\label{muKnown}
 \end{equation}
 for any BMSC $q_{Y|X}$ with capacity $\mathrm{C}(q_{Y|X})\in (0,1)$. We note from~\cite[Th.~48]{PPV10} (also \cite{Strassen} and~\cite{Hayashi09}) that the \emph{optimal} scaling exponents (optimized over all codes) are equal to~$2$ for $\varepsilon\in (0, 1/2)$ for non-degenerate DMCs.
 For a general BMC which does not need to be symmetric, we will see later in Lemma~\ref{lemmaBMCupperBound}, a stepping stone for establishing our main result, that the upper bound 4.714 in~$\eqref{muKnown}$ continues to hold. In this paper, we are interested in the scaling exponent of save-and-transmit polar codes for the binary-input EH channel, which is formally defined as follows.
\medskip
\begin{Definition} \label{defScalingExpPolar}
Fix an~$\varepsilon\in(0,1)$ and a binary-input EH channel $q_{Y|X}$ with capacity $\mathrm{C}(q_{Y|X};P)$ (defined in~\eqref{defCDMC}).
The \emph{scaling exponent for the binary-input EH channel restricted to save-and-transmit polar coding} is defined as
\begin{align*}
\mu_\varepsilon^{\text{\tiny PC-EH}} \triangleq \liminf_{N\rightarrow \infty} \inf \! \left\{\!\frac{-\log N}{\log \! \left|\mathrm{C}(q_{Y|X};P)- \frac{|\mathcal{I}|}{N}\right|}\!\left| \, \parbox[c]{1.3 in}{\small A save-and-transmit \\ $(m, (n, p_X,  \mathcal{I}, \lambda_{\mathcal{I}^c}), \varepsilon)$-EH polar code exists where $m+n=N$} \! \right.\right\}\! .
\end{align*}
\end{Definition}

The following theorem is the main result of this paper, which shows that 4.714, the upper bound on $\mu_\varepsilon^{\text{\tiny PC-BMC}}$ in~\eqref{muKnown} for BMSCs without cost constraints, remains to be a valid upper bound on the scaling exponent for the binary-input EH channel in spite of the additional EH constraints~\eqref{EHconstraintPolar}. The proof of the main result will be provided in Section~\ref{sectionProofofMainResult}.
\medskip
\begin{Theorem} \label{thmMainResult}
For any $\varepsilon\in(0,1)$ and any binary-input EH channel,
\begin{equation*}
\mu_\varepsilon^{\text{\tiny PC-EH}}\le 4.714.
\end{equation*}
\end{Theorem}

Theorem~\ref{thmMainResult} states that $4.714$ remains to be a valid upper bound on the scaling exponent of polar codes for the binary-input EH channel. This implies that the EH constraints do not worsen the rate of convergence to capacity if polar codes are employed. The chief intuition of this result is the following: We design the length of the saving phase~$m$ sufficiently small so that the convergence rate to the capacity $\mathrm{C}(q_{Y|X})$ is not affected. Yet, this choice of~$m$ ensures that the probability that the EH constraints are violated is small (cf.\ Proposition~\ref{propositionCharacteristicFunctionDMC}), and essentially does not significantly worsen the overall probability of decoding error. An auxiliary contribution of this paper is that the upper bound on the scaling exponent holds for binary-input memoryless asymmetric channels, which is established in Lemma~\ref{lemmaBMCupperBound} as an important step to proving Theorem~\ref{thmMainResult}.

\section{Proof of the Main Result} \label{sectionMainResultProof}
In this section, we will first analyze save-and-transmit EH-polar codes described in Definition~\ref{defPolarCodeEH} with randomized encoding functions $\lambda_{\mathcal{I}^c}$ for the frozen bits indexed by~$\mathcal{I}^c$. This randomized approach has been used in~\cite[Sec.~III-A]{HondaYamamoto13} for generalizing polarization results for symmetric channels to asymmetric channels, and it is also useful for analyzing save-and-transmit polar codes under the EH constraints~\eqref{EHconstraintPolar}. The proof of Theorem~\ref{thmMainResult} will be presented in Section~\ref{sectionProofofMainResult} after establishing two important lemmas concerning polar codes with randomized frozen bits in Section~\ref{sectionTwoImpLemma}.
\subsection{Polar Codes with Randomized Frozen Bits} \label{sectionTwoImpLemma}
Here we bound the difference between the code distribution of the EH-polar code and the one used to compute the Bhattacharyya parameters that appear in the code as described in Definition~\ref{defPolarCode}. Fix a $p_X$ and a $k\in\mathbb{N}$, and let $n= 2^k$. Define $p_{U^n, X^n, Y^n}$ as in~\eqref{defP}. In addition, for each set $\mathcal{A}\subseteq \{1, 2, \ldots, n\}$, define the set of $|\mathcal{A}|$-dimensional tuples of mappings
\begin{align*}
\Gamma(\mathcal{A})\triangleq\left\{(\lambda_i|\, i\in\mathcal{A})\left|\, \parbox[c]{2 in}{For each $i\in\mathcal{A}$, the domain and range of mapping $\lambda_i$ are $\{0,1\}^{i-1}$ and $\{0,1\}$ respectively}\right.\right\}.
\end{align*}
Construct a random variable $\Lambda_\mathcal{A}\triangleq (\Lambda_i|\, i\in\mathcal{A})$ distributed on $\Gamma(\mathcal{A})$ according to $r_{\Lambda_\mathcal{A}}$ such that
\begin{align}
r_{\Lambda_\mathcal{A}} = \prod_{i\in\mathcal{A}}r_{\Lambda_i} \label{defDistRprod}
\end{align}
and for all $i\in\mathcal{A}$ and all $u^{i-1}\in \{0,1\}^{i-1}$,
\begin{equation}
r_{\Lambda_i(u^{i-1})}(u_i) = p_{U_i|U^{i-1}}(u_i|u^{i-1}) \label{defDistRlambdaI}
\end{equation}
for each $u_i\in\{0,1\}$. Recalling the definitions of $\mathcal{I}$ and $\mathcal{I}^c$ in~\eqref{defInformationBitSet} and~\eqref{defFrozenBitSet} respectively, we consider an~$(n, p_X,  \mathcal{I}, \lambda_{\mathcal{I}^c})$-polar code for each $\lambda_{\mathcal{I}^c}\in \Gamma(\mathcal{I}^c)$. Let $r_{U^n, X^n, Y^n|\Lambda_{\mathcal{I}^c}=\lambda_{\mathcal{I}^c}}$ be the distribution induced by the $(n, p_X,  \mathcal{I}, \lambda_{\mathcal{I}^c})$-polar code, and let $r_{\Lambda_{\mathcal{I}^c}, U^n, X^n, Y^n}$ be the distribution induced by the randomized $(n, p_X,  \mathcal{I}, \Lambda_{\mathcal{I}^c})$-polar code where
\begin{align}
r_{\Lambda_{\mathcal{I}^c}, U^n, X^n, Y^n}(\lambda_{\mathcal{I}^c}, u^n, x^n, y^n)   \triangleq
r_{\Lambda_{\mathcal{I}^c}}(\lambda_{\mathcal{I}^c})r_{U^n, X^n, Y^n|\Lambda_{\mathcal{I}^c}=\lambda_{\mathcal{I}^c}}(u^n, x^n, y^n). \label{defDistR}
\end{align}
Then, we have the following lemma which characterizes the total variation distance between $r_{U^n, X^n, Y^n}$ defined in~\eqref{defDistR} and $p_{U^n, X^n, Y^n}$ defined in~\eqref{defP}. Since the proof of the lemma is similar to the proof of~\cite[Lemma~1]{HondaYamamoto13}, it is deferred to Appendix~\ref{appendixB}.
\medskip
\begin{Lemma} \label{lemmaVD}
For the randomized $(n, p_X,  \mathcal{I}, \Lambda_{\mathcal{I}^c})$-polar code where $\Lambda_{\mathcal{I}^c}\sim r_{\Lambda_{\mathcal{I}^c}}$, the total variation distance between $p_{U^n, X^n, Y^n}$ and $r_{U^n, X^n, Y^n}$ satisfies
\begin{equation}
\|p_{U^n, X^n, Y^n} - r_{U^n, X^n, Y^n}\| \le \frac{\sqrt{\ln 2}}{n}. \label{lemmaVDstatement}
\end{equation}
\end{Lemma}
\medskip

It has been shown in \cite[Th.~2]{MHU15} that $4.714$ is an upper bound on the scaling exponent for any
for any BMSC. The following lemma implies that $4.714$ is a valid upper bound on the scaling exponent for any BMC even if it is asymmetric, which serves as a stepping stone for the proof of Theorem~\ref{thmMainResult}. Although the proof of the following lemma is similar to the proof of~\cite[Th.~3]{HondaYamamoto13}, it is provided here to facilitate understanding.
\medskip
\begin{Lemma}\label{lemmaBMCupperBound}
Let $\mu = 4.714$ and fix a $p_X$ and a binary-input channel~$q_{Y|X}$. There exists a $t>0$ such that the following holds: For any $n$ which equals $2^k$ for some $k\in\mathbb{N}$, there exists a randomized~$(n, p_X,  \mathcal{I}_n, \Lambda_{\mathcal{I}^c}, \varepsilon_n)$-polar code with
\begin{equation*}
\frac{|\mathcal{I}_n|}{n} \ge I_{p_X q_{Y|X}}(X;Y)-  \frac{t}{n^{1/\mu}}
\end{equation*}
and
\begin{equation}
\varepsilon_n \le \frac{2\sqrt{\ln 2}}{n} + \frac{1}{2 n^3}. \label{lemmaBMupperBoundSt2}
\end{equation}
\end{Lemma}
\begin{IEEEproof}
Fix a binary-input channel $q_{Y|X}$ and a $p_X$, and define $p_{U^n, X^n, Y^n}$ as in~\eqref{defP} for each $n\in\mathbb{N}$. In addition, define
\begin{align}
\mathcal{I}_n\!\triangleq\! \left\{\parbox[c]{1 in}{$  i\in \{1, 2, \ldots, n\}$}\!\!\left|\parbox[c]{1.9 in}{$ Z_{p_{U^n, X^n, Y^n}}(U_i|U^{i-1}, Y^n) \!\le \! \frac{1}{n^4}, \\
 Z_{p_{U^n, X^n, Y^n}}(U_i|U^{i-1}) \ge 1-\frac{1}{n^4}$}  \!\!\!\right.  \right\} \label{defIn}
 \end{align}
for each $n\in\mathbb{N}$. By Lemma~\ref{lemmaPolar}, there exists a $t>0$ such that for each~$n$ which equals $2^k$ for some $k\in\mathbb{N}$,
\begin{equation}
\frac{|\mathcal{I}_n|}{n}\ge I_{p_X q_{Y|X}}(X; Y) - \frac{t}{n^{1/\mu}}. \label{lemmaBMupperBoundEq0}
\end{equation}
It remains to prove~\eqref{lemmaBMupperBoundSt2}. To this end, we fix $n$ and let $r_{\Lambda_{\mathcal{I}_n^c}, U^n, X^n, Y^n}$ be the distribution induced by the randomized $(n, p_X, \mathcal{I}_n, \Lambda_{\mathcal{I}_n^c})$-polar code, where $r_{\Lambda_{\mathcal{I}_n^c}, U^n, X^n, Y^n}$ is as defined in~\eqref{defDistR}. For the randomized $(n, p_X, \mathcal{I}_n, \Lambda_{\mathcal{I}_n^c})$-polar code, let $\varphi:\mathcal{Y}^n\rightarrow \mathcal{U}^n$ characterize the overall decoding function induced by the successive cancellation decoders $\varphi_i$'s (cf.\ Definition~\ref{defPolarCode}) such that $\hat U^n = \varphi(Y^n)$ is the output of the decoders given the channel output~$Y^n$, and consider the following probability of decoding error:
\begin{align}
\Pr_{r_{\Lambda_{\mathcal{I}_n^c}, U^n, X^n, Y^n}}\left\{U^n \ne \varphi(Y^n)\right\} 
& = \sum_{(u^n, y^n)\in \mathcal{U}^n\times \mathcal{Y}^n}r_{U^n, Y^n}(u^n, y^n) \mathbf{1}\{u^n \ne \varphi(y^n)\}\notag\\
& \stackrel{\text{(a)}}{\le} 2\|r_{U^n, Y^n} - p_{U^n, Y^n}\|  + \hspace{-.5 in} \sum_{\qquad \qquad (u^n, y^n)\in \mathcal{U}^n\times \mathcal{Y}^n}\hspace{-.5 in}  p_{U^n, Y^n}(u^n, y^n) \mathbf{1}\{u^n \ne \varphi(y^n)\}\notag\\
& \stackrel{\text{(b)}}{\le}\frac{2\sqrt{\ln 2}}{n} + \hspace{-.5 in}\sum_{\qquad \qquad(u^n, y^n)\in \mathcal{U}^n\times \mathcal{Y}^n}\hspace{-.5 in}p_{U^n, Y^n}(u^n, y^n) \mathbf{1}\{u^n \ne \varphi(y^n)\}\notag\\
& \stackrel{\text{(c)}}{\le} \frac{2\sqrt{\ln 2}}{n} + \sum_{i=1}^n \hspace{-.5 in}\sum_{\qquad \qquad(u^i, y^n)\in \mathcal{U}^i\times \mathcal{Y}^n}\hspace{-.5 in}p_{U^i, Y^n}(u^i, y^n) \mathbf{1}\{u_i \ne \varphi_i(u^{i-1},y^n)\}\notag\\
&\stackrel{\text{(d)}}{=}\frac{2\sqrt{\ln 2}}{n} +  \sum_{ i\in \mathcal{I}_n}\!\! \hspace{-.5 in} \sum_{\qquad \qquad(u^i, y^n)\in \mathcal{U}^i\times \mathcal{Y}^n}\hspace{-.5 in}p_{U^i, Y^n}(u^i, y^n) \mathbf{1}\{u_i \ne \varphi_i(u^{i-1},y^n)\} \label{lemmaBMupperBoundEq1}
\end{align}
where
\begin{enumerate}
\item[(a)] follows from the triangle inequality.
\item[(b)] follows from Lemma~\ref{lemmaVD}.
\item[(c)] follows from the definition of the successive cancellation decoders in Definition~\ref{defPolarCode} and the fact that $\{u^n \ne \varphi(y^n)\}$ can be written as a union of disjoint events as
    \begin{align*}
\{u^n \ne \varphi(y^n)\}=
   \bigcup_{i=1}^n \left\{\{u_i \ne \varphi_i(\hat u^{i-1}, y^n) \}\cap \{u^{i-1}=\hat u^{i-1}\}\right\}.
    \end{align*}
    \item[(d)] follows from the fact due to Definition~\ref{defPolarCode} that for all $i\in\mathcal{I}_n^c$,
        \begin{equation*}
        \Pr_{p_{U^i, Y^n}}\{U_i \ne \varphi_i(U^{i-1},Y^n)\}=0.
        \end{equation*}
\end{enumerate}
Consider the following chain of inequalities for each $i\in \mathcal{I}_n$:
\begin{align}
& \sum_{(u^i, y^n)\in \mathcal{U}^i\times \mathcal{Y}^n}p_{U^i, Y^n}(u^i, y^n) \mathbf{1}\{u_i \ne \varphi_i(u^{i-1},y^n)\} \notag\\*
&  = \hspace{-0.6 in} \sum_{\qquad \qquad(u^{i-1}, y^n)\in \mathcal{U}^{i-1}\times \mathcal{Y}^n}\hspace{-0.6 in}p_{U^{i-1}, Y^n}(u^{i-1}, y^n)\sum_{u_i\in\mathcal{U}}p_{U_i|U^{i-1}, Y^n}(u_i|u^{i-1}, y^n)\mathbf{1}\{u_i \ne \varphi_i(u^{i-1},y^n)\}\notag\\
&\stackrel{\eqref{defSCdecoder}}{\le} \hspace{-0.6 in} \sum_{\qquad \qquad(u^{i-1}, y^n)\in \mathcal{U}^{i-1}\times \mathcal{Y}^n}\hspace{-0.6 in}p_{U^{i-1}, Y^n}(u^{i-1}, y^n)\sum_{u_i\in\mathcal{U}}p_{U_i|U^{i-1}, Y^n}(u_i|u^{i-1}, y^n)\sqrt{\frac{p_{U_i|U^{i-1}, Y^n}(u_i + 1|u^{i-1}, y^n)}{p_{U_i|U^{i-1}, Y^n}(u_i|u^{i-1}, y^n)}}\notag\\
&\stackrel{\eqref{defBhattacharyya}}{=} Z_{p_{U^i,Y^n}}(U_i|U^{i-1},Y^n)/2\notag \\
&\stackrel{\eqref{defIn}}{\le}\frac{1}{2n^4}.  \label{lemmaBMupperBoundEq2}
\end{align}
Combining \eqref{lemmaBMupperBoundEq1} and \eqref{lemmaBMupperBoundEq2}, we obtain
\begin{align}
\Pr_{r_{\Lambda_{\mathcal{I}_n^c}, U^n, X^n, Y^n}}\left\{U^n \ne \varphi(Y^n)\right\} \le \frac{2\sqrt{\ln 2}}{n} + \frac{1}{2n^3}. \label{lemmaBMupperBoundEq3}
\end{align}
The lemma then follows from~\eqref{lemmaBMupperBoundEq0} and~\eqref{lemmaBMupperBoundEq3}.
\end{IEEEproof}

\subsection{Save-and-Transmit EH-Polar Codes with Randomized Frozen Bits}\label{sectionProofofMainResult}
In this section, we will use the randomized polar codes defined in the previous section to construct save-and-transmit EH-polar codes and establish the following theorem, which will immediately lead to Theorem~\ref{thmMainResult}.
\medskip
\begin{Theorem} \label{thmSaveAndTransmit}
Let $\mu = 4.714$ and fix a binary-input EH channel~$q_{Y|X}$ and $p_X$ such that
\begin{equation}
\E_{p_X}[X]=\E_{p_{E_1}}[E_1]=P. \label{thmSaveAndTransmitSt1}
 \end{equation}
 Define $a\triangleq \max\left\{\E_{p_{E_1}}[E_1^2], e \right\}$ as in~\eqref{propositionCharFuncAssump2DMC}. Then, there exists a $t>0$ such that the following holds: For any $n\ge 3$ which equals $2^k$ for some $k\in\mathbb{N}$ and sufficiently large such that
 \begin{equation}
 \frac{n}{\ln n}\ge \frac{a}{P^2}, \label{thmSaveAndTransmitSt2}  \end{equation}
  there exists a save-and-transmit~$(m, (n, p_X,  \mathcal{I}_n, \Lambda_{\mathcal{I}^c}), \varepsilon_n)$-EH polar code with
\begin{equation*}
m \le \frac{6\sqrt{an\ln n}}{P}  +1,
\end{equation*}
\begin{equation*}
\frac{|\mathcal{I}_n|}{n} \ge I_{p_X q_{Y|X}}(X;Y)-  \frac{t}{n^{1/\mu}}
\end{equation*}
and
\begin{equation*}
\varepsilon_n \le \frac{e^{0.4} }{n\ln n} + \frac{4\sqrt{\ln 2}}{n} + \frac{1}{2 n^3}.
\end{equation*}
\end{Theorem}
\begin{IEEEproof}
Fix a binary-input EH channel~$q_{Y|X}$ and $p_X$ such that~\eqref{thmSaveAndTransmitSt1} holds. By Lemma~\ref{lemmaBMCupperBound}, there exists a~$t>0$ such that the following holds: For any $n$ which equals $2^k$ for some $k\in\mathbb{N}$, there exists a randomized~$(n, p_X,  \mathcal{I}_n, \Lambda_{\mathcal{I}^c}, \delta_n)$-polar code with
\begin{equation}
\frac{|\mathcal{I}_n|}{n} \ge I_{p_X q_{Y|X}}(X;Y)-  \frac{t}{n^{1/\mu}} \label{defINinProof}
\end{equation}
and
\begin{equation}
\delta_n \le \frac{2\sqrt{\ln 2}}{n} + \frac{1}{2 n^3}. \label{defDeltaN}
\end{equation}
Define \begin{equation}
m \triangleq \left\lceil\frac{6\sqrt{an\ln n}}{P}  \right\rceil \label{defM}
\end{equation}
for each $n\in\mathbb{N}$. Fix a sufficiently large~$n\ge 3$ that satisfies~\eqref{thmSaveAndTransmitSt2} and consider the corresponding save-and-transmit $(m, (n, p_X,  \mathcal{I}_n, \Lambda_{\mathcal{I}^c}))$-EH code as described in Definition~\ref{defPolarCodeEH} where the $(n, p_X,  \mathcal{I}_n, \Lambda_{\mathcal{I}^c}, \delta_n)$-polar code with stochastic functions $\Lambda_{\mathcal{I}^c}$ serves as an effective code of the save-and-transmit $(m, (n, p_X,  \mathcal{I}_n, \Lambda_{\mathcal{I}^c}))$-EH code. Let $N\triangleq m+n$, and let $r_{E^N ,U^n, X^N, Y^N, \hat U^n}$ be the distribution induced by the save-and-transmit $(m, (n, p_X,  \mathcal{I}_n, \Lambda_{\mathcal{I}^c}))$-EH code which satisfies the EH constraints~\eqref{EHconstraintPolar}, where $U^n$ denotes the information and frozen bits chosen by the effective code and $\hat U^n$ denote the estimate of~$U^n$ declared by node~$\mathrm{d}$ (cf.\ Definition~\ref{defPolarCodeEH}). Using~\eqref{memorylessStatement}, we have
\begin{align}
 r_{E^N ,U^n, X^N, Y^N, \hat U^n} =  r_{U^n} \left(\prod_{i=1}^N p_{E_i} r_{X_i|U^n, E^i} r_{Y_i|X_i}\right)r_{\hat U^n|Y^N} \label{defRinMainProofFormer}
\end{align}
where $r_{Y_i|X_i}(y_i|x_i)= q_{Y|X}(y_i|x_i)$ for all $i\in\{1, 2, \ldots, N\}$, all $x_i\in\mathcal{X}$ and all $y_i\in\mathcal{Y}$. In addition, let $\tilde X^n$ be the transmitted codeword induced by the randomized~$(n, p_X,  \mathcal{I}_n, \Lambda_{\mathcal{I}^c}, \delta_n)$-polar code when there is no cost constraint, and define
\begin{align}
 r_{E^N ,U^n, X^N, Y^N, \hat U^n, \tilde X^n}\triangleq  r_{E^N ,U^n, X^N, Y^N, \hat U^n} r_{\tilde X^n|U^n} \label{defRinMainProof}
\end{align}
where $r_{\tilde X^n|U^n}$ characterizes the inverse polarization mapping used by the $(n, p_X,  \mathcal{I}_n, \Lambda_{\mathcal{I}^c}, \delta_n)$-polar code according to~\eqref{polarizationInvMap}. In the rest of the proof, all the probability terms are evaluated according to $r_{E^N ,U^n, X^N, Y^N, \hat U^n, \tilde X^n}$ unless specified otherwise. The probability of decoding error of the save-and-transmit $(m,  (n, p_X,  \mathcal{I}_n, \Lambda_{\mathcal{I}^c}))$-EH code can be bounded as
\begin{align}
\Pr\left\{U^n \ne \hat U^n \right\} 
& \le \Pr\left\{\parbox[c]{2.4 in}{$\{U^n \ne \hat U^n \}\cap \bigcap\limits_{i=m+1}^{m+n} \left\{\sum\limits_{\ell=1}^{i-m}  \tilde X_{\ell} \le \sum\limits_{\ell=1}^i E_\ell \right\}$}\right\}+ \Pr\left\{ \bigcup_{i=m+1}^{m+n} \left\{\sum_{\ell=1}^{i-m} \tilde X_{\ell} > \sum_{\ell=1}^i E_\ell \right\}\right\}. \label{eqn1MainProof}
\end{align}
Consider
\begin{align}
\Pr\left\{\parbox[c]{2.4 in}{$\{U^n \ne \hat U^n \}\cap  \bigcap\limits_{i=m+1}^{m+n} \left\{\sum\limits_{\ell=1}^{i-m}  \tilde X_{\ell} \le \sum\limits_{\ell=1}^i E_\ell \right\}$}\right\} 
& \stackrel{\eqref{defFiPolar}}{=} \Pr\left\{\!\parbox[c]{2.4 in}{$\{U^n \ne \hat U^n \}\cap \bigcap\limits_{i=m+1}^{m+n} \left\{\sum\limits_{\ell=1}^{i-m}  \tilde X_{\ell} \le \sum\limits_{\ell=1}^i E_\ell \right\} \\ \text{ } \cap \left\{\parbox[c]{2 in}{$(X_{m+1}, X_{m+2}, \ldots,X_{m+n})=\tilde X^n$}\right\}$}\right\} \notag\\*
&\le \Pr\left\{\parbox[c]{3 in}{$\{U^n \ne \hat U^n \}\cap  \left\{\parbox[c]{1.96 in}{$(X_{m+1}, X_{m+2}, \ldots,X_{m+n})=\tilde X^n$}\right\}$}\!\!\right\}\!.\label{eqn1MainProof*}
\end{align}
By inspecting \eqref{defpUgivenXinPolar} in Definition~\ref{defPolarCode}, \eqref{polarizationInvMap} and~\eqref{defSCdecoderEH} in Definition~\ref{defPolarCodeEH} and the definition of~$r$ in~\eqref{defRinMainProof}, we conclude that the upper bound in~\eqref{eqn1MainProof*} cannot exceed the probability of decoding error of the effective code, which implies that
\begin{align}
\Pr\left\{\parbox[c]{2.4 in}{$\{U^n \ne \hat U^n \}\cap  \bigcap\limits_{i=m+1}^{m+n} \left\{\sum\limits_{\ell=1}^{i-m}  \tilde X_{\ell} \le \sum\limits_{\ell=1}^i E_\ell \right\}$}\right\}  \le \delta_n. \label{eqn2MainProof}
\end{align}
In order to bound the second probability in~\eqref{eqn1MainProof}, recall that $p_{X^n}=\prod_{i=1}^n p_{X_i}$ and consider the following chain of inequalities:
\begin{align}
&\Pr\left\{ \bigcup_{i=m+1}^{m+n} \left\{\sum_{\ell=1}^{i-m}\tilde  X_{\ell} > \sum_{\ell=1}^i E_\ell \right\}\right\} \notag\\*
& \stackrel{\text{(a)}}{=} \Pr_{r_{E^N} r_{\tilde X^n}}\left\{ \bigcup_{i=m+1}^{m+n} \left\{\sum_{\ell=1}^{i-m}\tilde  X_{\ell} > \sum_{\ell=1}^i E_\ell \right\}\right\} \notag \\
&  = \int_{\mathbb{R}_+^{m+n}} \sum_{\tilde x^n\in \{0,1\}^n} r_{E^{m+n}}(e^{m+n})r_{\tilde X^n}(\tilde x^n) \mathbf{1}\left\{  \bigcup_{i=m+1}^{m+n} \left\{\sum_{\ell=1}^{i-m}\tilde  x_{\ell} > \sum_{\ell=1}^i e_\ell \right\} \right\} \mathrm{d} e^{m+n} \notag\\
& \le 2\|r_{\tilde X^n} - p_{X^n}\| + \int_{\mathbb{R}_+^{m+n}} \sum_{x^n\in \{0,1\}^n} r_{E^{m+n}}(e^{m+n})p_{X^n}(x^n) \mathbf{1}\left\{  \bigcup_{i=m+1}^{m+n} \left\{\sum_{\ell=1}^{i-m} x_{\ell} > \sum_{\ell=1}^i e_\ell \right\} \right\} \mathrm{d} e^{m+n} \notag\\
& \stackrel{\text{(b)}}{\le}\frac{2\sqrt{\ln 2}}{n} +  \Pr_{r_{E^{m+n}} p_{X^n}}\left\{ \bigcup_{i=m+1}^{m+n} \left\{\sum_{\ell=1}^{i-m} X_{\ell} > \sum_{\ell=1}^i E_\ell \right\}\right\}  \label{eqn3MainProof}
\end{align}
where
\begin{enumerate}
\item[(a)] follows from~\eqref{defRinMainProofFormer} and~\eqref{defRinMainProof}.
\item[(b)] follows from Lemma~\ref{lemmaVD}.
\end{enumerate}
Since $r_{E^{m+n}} = \prod_{i=1}^{m+n}p_{E_i}$ by \eqref{defRinMainProofFormer} and $p_{X^n}=\prod_{i=1}^n p_{X_i}$, it follows from Proposition~\ref{propositionCharacteristicFunctionDMC} and~\eqref{defM} that
\begin{align*}
\Pr_{r_{E^{m+n}} p_{X^n}}\left\{ \bigcup_{i=m+1}^{m+n} \left\{\sum_{\ell=1}^{i-m} X_{\ell} > \sum_{\ell=1}^i E_\ell \right\}\right\}\le \frac{e^{0.4}}{n\ln n},
\end{align*}
which implies from \eqref{eqn3MainProof} that
\begin{align}
\Pr\left\{ \bigcup_{i=m+1}^{m+n} \left\{\sum_{\ell=1}^{i-m}\tilde  X_{\ell} > \sum_{\ell=1}^i E_\ell \right\}\right\}\le \frac{2\sqrt{\ln 2}}{n} + \frac{e^{0.4}}{n\ln n}.  \label{eqn4MainProof}
\end{align}
Combining \eqref{defDeltaN}, \eqref{eqn1MainProof}, \eqref{eqn2MainProof} and~\eqref{eqn4MainProof}, we conclude that the probability of decoding error $\varepsilon_n$ of the save-and-transmit $(m, (n, p_X,  \mathcal{I}_n, \Lambda_{\mathcal{I}^c}))$-EH polar code satisfies
\begin{equation}
 \varepsilon_n \le \frac{e^{0.4} }{n\ln n} + \frac{4\sqrt{\ln 2}}{n} + \frac{1}{2 n^3}. \label{eqn5MainProof}
\end{equation}
Consequently, the theorem follows from the fact that the save-and-transmit $(m, (n, p_X,  \mathcal{I}_n, \Lambda_{\mathcal{I}^c}))$-EH polar code satisfies~\eqref{defINinProof}, \eqref{defM} and \eqref{eqn5MainProof} for each sufficiently large~$n\ge 3$ that satisfies~\eqref{thmSaveAndTransmitSt2}.
\end{IEEEproof}
\medskip

We are ready to present the proof of Theorem~\ref{thmMainResult}.
\medskip
\begin{IEEEproof}[Proof of Theorem~\ref{thmMainResult}]
Choose a $p_X^*$ such that $\E_{p_X^*}[X]=P$ and
\begin{align}
I_{p_X^* q_{Y|X}}(X;Y)&=\max\limits_{p_X: \E_{p_X}[X]=P}I_{p_X q_{Y|X}}(X;Y)\notag\\*
&\stackrel{\eqref{defCDMC}}{=}\mathrm{C}(q_{Y|X};P). \label{thmFinalProofEq1}
\end{align}
Theorem~\ref{thmSaveAndTransmit} implies that there exist $\alpha_1>0$, $\alpha_2>0$ and $\alpha_3>0$ such that for all sufficiently large~$k$, a save-and-transmit $(m, (n, p_X,  \mathcal{I}_n, \Lambda_{\mathcal{I}^c}), \varepsilon_n)$-EH polar code exists where $n=2^k$,
\begin{equation}
m \le \alpha_1\sqrt{n\ln n}, \label{thmFinalProofEq2}
\end{equation}
\begin{equation}
\frac{|\mathcal{I}_n|}{n} \ge I_{p_X^* q_{Y|X}}(X;Y)-  \frac{\alpha_2}{n^{1/\mu}} \label{thmFinalProofEq3}
\end{equation}
and
\begin{equation}
\varepsilon_n \le \frac{\alpha_3}{n}. \label{thmFinalProofEq4}
\end{equation}
In addition, for all sufficiently large~$n$, we have
\begin{align}
m\stackrel{\eqref{thmFinalProofEq2}}{\le} n, \label{thmFinalProofEq4**}
\end{align}
\begin{align}
n^{1/\mu} > \alpha_1+\alpha_2 \label{thmFinalProofEq4***}
\end{align}
and
\begin{align}
\frac{m}{m+n}&\le \frac{m}{n}\notag\\
 &\stackrel{\eqref{thmFinalProofEq2}}{\le} \alpha_1\sqrt{\frac{\ln n}{n}} \notag\\
& \le \frac{\alpha_1}{n^{1/4.714}}\notag\\
&= \frac{\alpha_1}{n^{1/\mu}}. \label{thmFinalProofEq4*}
\end{align}
For such a save-and-transmit $(m, (n, p_X,  \mathcal{I}_n, \Lambda_{\mathcal{I}^c}), \varepsilon_n)$-EH polar code, we have for sufficiently large~$n$
\begin{align}
\frac{-\log (m+n)}{\log \left|\mathrm{C}(q_{Y|X};P)- \frac{|\mathcal{I}_n|}{m+n}\right|}
&\stackrel{\eqref{thmFinalProofEq1}}{=} \frac{-\log (m+n)}{\log \left|I_{p_X^* q_{Y|X}}(X;Y)- \frac{|\mathcal{I}_n|}{m+n}\right|}\notag\\
& =  \frac{\log (m+n)}{\log \left(1\big/\big|I_{p_X^* q_{Y|X}}(X;Y)- \frac{|\mathcal{I}_n|}{m+n}\big|\right)}\notag\\
& \stackrel{\text{(a)}}{\le} \frac{\log (m+n)}{\log\left|\frac{n^{1/\mu}}{\alpha_1+\alpha_2}\right| }\notag\\
&\stackrel{\eqref{thmFinalProofEq4***}}{=}  \frac{\log (m+n)}{\log\left(\frac{n^{1/\mu}}{\alpha_1+\alpha_2}\right) }\notag\\*
&\stackrel{\eqref{thmFinalProofEq4**}}{\le} \frac{\log (2n)}{ \frac{1}{\mu}\log n-\log (\alpha_1+\alpha_2)}  \label{thmFinalProofEq5}
\end{align}
where (a) follows from the fact that for sufficiently large~$n$, we have
\begin{align*}
I_{p_X^* q_{Y|X}}(X;Y)- \frac{|\mathcal{I}_n|}{m+n} 
&= I_{p_X^* q_{Y|X}}(X;Y)- \frac{|\mathcal{I}_n|}{n} + \frac{m|\mathcal{I}_n|}{n(m+n)} \notag\\
&\le I_{p_X^* q_{Y|X}}(X;Y)- \frac{|\mathcal{I}_n|}{n} + \frac{m}{m+n}\notag\\
&\stackrel{\eqref{thmFinalProofEq4*}}{\le} I_{p_X^* q_{Y|X}}(X;Y)- \frac{|\mathcal{I}_n|}{n} +  \frac{\alpha_1}{n^{1/\mu}}\notag\\
& \stackrel{\eqref{thmFinalProofEq3}}{\le} \frac{\alpha_1 + \alpha_2}{n^{1/\mu}}. \notag
\end{align*}
Since $\lim_{k\rightarrow\infty}\varepsilon_{2^k}=0$ by \eqref{thmFinalProofEq4}, it follows from~\eqref{thmFinalProofEq5} that for each $\varepsilon \in (0,1)$, there exists for each sufficiently large~$k$ a save-and-transmit $(m, (2^k, p_X,  \mathcal{I}_{2^k}, \Lambda_{\mathcal{I}^c}), \varepsilon)$-EH polar code such that
\begin{equation*}
\frac{-\log (m+2^k)}{\log \left|\mathrm{C}(q_{Y|X};P)- \frac{|\mathcal{I}_{2^k}|}{m+2^k}\right|} \le \frac{k+1}{ \frac{k}{\mu}-\log (\alpha_1+\alpha_2)},
\end{equation*}
which implies from Definition~\ref{defScalingExpPolar} that
\begin{equation*}
\mu_\varepsilon^{\text{\tiny PC-EH}}  \le \mu
\end{equation*}
for each $\varepsilon\in(0,1)$.
\end{IEEEproof}

 \section{Concluding Remarks} \label{sec:conclusion}
%
The encoding and decoding complexities of our proposed save-and-transmit polar codes are the same as that of the polar codes proposed for asymmetric channels in~\cite{HondaYamamoto13}. Therefore as discussed in~\cite[Sec.~III-B]{HondaYamamoto13}, the encoding and decoding complexities of our proposed save-and-transmit polar codes are at most $O(n \log n)$ as long as we allow pseudorandom numbers to be shared between the encoder and the decoder for encoding and decoding the randomized frozen bits. By a standard probabilistic argument, there must exist a deterministic encoder for the frozen bits such that the decoding error of the save-and-transmit polar code with the deterministic encoder is no worse than the polar code with randomized frozen bits. In the future, it may be fruitful to develop low-complexity algorithms for finding a good deterministic encoder for encoding the frozen bits. Other directions for future work can include exploring polar codes for EH channels under other asymptotic regimes such as the error exponent, moderate deviations or error floors regimes studied by Mondelli, Hassani and Urbanke~\cite{MHU15}.

\appendices
\section{Proof of Lemma~\ref{lemmaPolar}} \label{appendixA}
The proof of Lemma~\ref{lemmaPolar} relies on the following three propositions. The proof of Lemma~\ref{lemmaPolar} will be presented after stating the three propositions.

Before stating the first proposition, we define $s_X$ to be the uniform distribution on $\mathcal{X}$, define $s_{X^n}$ to be the distribution of~$n$ independent copies of~$X\sim s_X$, i.e., $s_{X^n}(x^n) = \prod_{i=1}^n s_{X}(x_i)$ for all $x^n\in \mathcal{X}^n$, and define
   \begin{equation}
  s_{U^n, X^n, Y^n}\triangleq s_{X^n} p_{U^n|X^n}\prod_{i=1}^n p_{Y_i|X_i} \label{defS}
  \end{equation}
  where $p_{U^n|X^n}$ characterizes the relation between $U^n$ and $X^n$ in~\eqref{defpUgivenX}.
   \begin{Proposition}[{\cite[Proposition~2]{arikan09}}] \label{propositionErrorPolar}
Fix a BMSC $q_{Y|X}$, a $k\in\mathbb{N}$ and an index set $\mathcal{I} \subseteq \{1, 2, \ldots, 2^k\}$. Let $n= 2^k$. Then, there exists an $(n, 2^{|\mathcal{I}|}, \varepsilon_n)$-code such that
 \begin{align*}
\varepsilon_n \le \sum_{i\in\mathcal{I}} Z_{s_{U^n, X^n, Y^n}}(U_i|U^{i-1}, Y^n) 
 \end{align*}
 where $s_{U^n, X^n, Y^n}$ is as defined in~\eqref{defS}.
  \end{Proposition}

    The following proposition can be derived in a straightforward manner from the proofs of~\cite[Ths.~1 and~2]{MHU15} and~\cite[Remark~4]{MHU15}.
  \medskip
  \begin{Proposition} \label{propositionPolar1}
 Fix a BMSC $q_{Y|X}$ and let $\mu=4.714$. Then, there exist two positive numbers~$t_1$ and~$t_2$ which do not depend on~$n$ such that for any $k\in\mathbb{N}$ and $n\triangleq 2^k$, we have
 \begin{align}
 \frac{1}{n}\left|\left\{ i\in\{1, 2, \ldots, n\}\left| Z_{s_{U^n, X^n, Y^n}}(U_i|U^{i-1}, Y^n) \le \frac{1}{n^4}  \right.  \right\}\right| \ge I_{s_X q_{Y|X}}(X; Y) - \frac{t_1}{n^{1/\mu}}. \label{st1InPropositionPolar}
 \end{align}
In addition, if
\begin{equation}
I_{s_X q_{Y|X}}(X; Y) = \max_{p_X}I_{p_X q_{Y|X}}(X;Y), \label{st1*InPropositionPolar}
\end{equation}
 then
  \begin{align}
 \frac{1}{n}\left|\left\{ i\in\{1, 2, \ldots, n\}\!\left|  Z_{s_{U^n, X^n, Y^n}}(U_i|U^{i-1}\!, Y^n) \ge 1- \! \frac{1}{n^4}  \right. \! \right\}\right| \ge 1- I_{s_X q_{Y|X}}(X; Y) - \frac{t_2}{n^{1/\mu}}. \label{st2InPropositionPolar}
 \end{align}
  \end{Proposition}
  \begin{IEEEproof}
  It follows from the proof of \cite[Th.~2]{MHU15} that there exists a mapping $h: [0, 1]\rightarrow [0,1]$ such that $h(0)=h(1)=0$, $h(x)>0$ for any $x\in(0,1)$ and
  \begin{equation}
  \sup\limits_{x\in(0,1), y\in [x\sqrt{2-x^2}, 2x-x^2]}\frac{h(x^2)+h(y)}{2h(x)} \le \frac{1}{2^{1/\mu}}.  \label{propositionPolarProofEqn1}
  \end{equation}
  Then, \eqref{st1InPropositionPolar} follows from the inequality in~\eqref{propositionPolarProofEqn1},~\cite[Eq.~(34) in proof of~Th.~1]{MHU15} and~\cite[Remark~4]{MHU15}. It remains to prove \eqref{st2InPropositionPolar}. To this end, suppose~\eqref{st1*InPropositionPolar} holds. Define
  \begin{align*}
  \mathcal{I}\triangleq \left\{ i\in\{1, 2, \ldots, n\}\left| Z_{s_{U^n, X^n, Y^n}}(U_i|U^{i-1}, Y^n) \le \frac{1}{n^4}  \right.  \right\},
  \end{align*}
which implies from Proposition~\ref{propositionErrorPolar} that there exists an $(n, 2^{|\mathcal{I}|}, n^{-3})$-code. Since the capacity of the channel is equal to $I_{s_{X}q_{Y|X}}(X; Y)$ by~\eqref{st1*InPropositionPolar}, it follows from \cite[Th.~48]{PPV10} (also \cite{Strassen} and~\cite{Hayashi09}) that there exists a $\lambda_1 > 0$ such that
\begin{align*}
\log 2^{|\mathcal{I}|} \le n I_{s_X q_{Y|X}}(X; Y) + \lambda_1\sqrt{n},
\end{align*}
which implies that
\begin{align}
\frac{|\mathcal{I}| }{n}\le I_{s_X q_{Y|X}}(X; Y) + \frac{\lambda_1}{\sqrt{n}}. \label{propositionPolarProofEqn2}
\end{align}

On the other hand, define
 \begin{align*}
  \mathcal{J}\triangleq \left\{ i\in\{1, 2, \ldots, n\}\left| \text{$Z_{s_{U^n, X^n, Y^n}}(U_i|U^{i-1}, Y^n)\in \left[\frac{1}{n^4}, 1- \frac{1}{n^4}\right] $} \right.  \right\}.
  \end{align*}
  It has been shown in~\cite[Eq.~(65) and Remark~4]{MHU15} that there exists a $\lambda_2 >0$ such that
  \begin{align}
  \frac{|\mathcal{J}|}{n} \le \frac{\lambda_2}{n^{1/\mu}}. \label{propositionPolarProofEqn3}
  \end{align}
Statement~\eqref{st2InPropositionPolar} then follows from~\eqref{propositionPolarProofEqn2} and~\eqref{propositionPolarProofEqn3}.
  \end{IEEEproof}
\medskip

 The following construction of $\hat p_{\hat U^n, \hat X^n, \hat Y^n}$ and the subsequent proposition are the main tools used in~\cite{HondaYamamoto13} for generalizing polarization results for symmetric channels to asymmetric channels. Fix any distribution $p_X$ defined on~$\mathcal{X}=\{0,1\}$. We define $\hat p_{\hat U^n, \hat X^n, \hat Y^n}$ based on~$p_X$ in several steps as follows. Define $\hat p_{\hat X}$ to be the uniform distribution over $\hat{\mathcal{X}}\triangleq \{0,1\}$, define $\hat{\mathcal{Y}}\triangleq \{0,1\}\times \mathcal{Y}$, define $\hat q_{\hat Y| \hat X}$ such that
\begin{equation}
\hat q_{\hat Y|\hat X}((\hat x + x,y)|\hat x) = p_X(x)q_{Y|X}(y|x) \label{defHatQ}
\end{equation}
for all $(\hat x, x, y)\in \hat{\mathcal{X}}\times \mathcal{X}\times \mathcal{Y}$ where $+$ denotes addition over $\mathrm{GF}(2)$, define $\hat p_{\hat X^n, \hat Y^n}$ such that
\begin{align}
\hat p_{\hat X^n, \hat Y^n}(\hat x^n,(\hat x^n + x^n, y^n))= \prod_{i=1}^n \hat p_{\hat X}(\hat x_i) \hat q_{\hat Y|\hat X}((\hat x_i + x_i,y_i)|\hat x_i) \label{defHatPXY}
\end{align}
for all $(\hat x^n, x^n, y^n)\in \hat{\mathcal{X}}^n\times \mathcal{X}^n\times \mathcal{Y}^n$, and define $\hat p_{\hat U^n, \hat X^n, \hat Y^n}$ such that
\begin{align}
\hat p_{\hat U^n, \hat X^n, \hat Y^n}(\hat u^n, \hat x^n, (\hat x^n + x^n, y^n))
&\triangleq \hat p_{\hat X^n, \hat Y^n}(\hat x^n, (\hat x^n + x^n, y^n))p_{U^n|X^n}(\hat u^n|\hat x^n) \label{defHatP} \\*
& \stackrel{\eqref{defHatPXY}}{=}p_{U^n|X^n}(\hat u^n|\hat x^n) \prod_{i=1}^n \hat p_{\hat X}(\hat x_i) \hat q_{\hat Y|\hat X}((\hat x_i + x_i,y_i)|\hat x_i) \notag\\*
&\stackrel{\eqref{defHatQ}}{=} p_{U^n|X^n}(\hat u^n|\hat x^n) \prod_{i=1}^n \left( \hat p_{\hat X}(\hat x_i) p_X(x_i)q_{Y|X}(y_i|x_i)\right) \notag
\end{align}
for all $(\hat x^n, x^n, y^n)\in \hat{\mathcal{X}}^n\times \mathcal{X}^n\times \mathcal{Y}^n$, where $p_{U^n|X^n}$ was defined in~\eqref{defpUgivenX}.
\medskip
\begin{Proposition}[{\cite[Th.~2]{HondaYamamoto13}}] \label{propositionPolar2}
For any binary-input channel $q_{Y|X}$ and any $p_X$, define $p_{U^n, X^n, Y^n}$ and $\hat p_{\hat U^n, \hat X^n, \hat Y^n}$ as in~\eqref{defP} and~\eqref{defHatP} respectively. Then, the following equations hold for each $i\in\{1, 2, \ldots, n\}$:
\begin{equation*}
p_{U^i, Y^n}(u^i, y^n) = 2^{n-1} \hat p_{\hat U^{i-1}, \hat Y^n| \hat U_i}(u^{i-1}, (0^n, y^n)|u_i)
\end{equation*}
for each $(u^i, y^n)\in \mathcal{U}^i\times \mathcal{Y}^n$ where $0^n$ denotes the $n$-dimensional zero tuple,
and
\begin{align*}
Z_{p_{U^n, X^n, Y^n}}(U_i|U^{i-1}, Y^n) = Z_{\hat p_{\hat U^n, \hat X^n, \hat Y^n}}(\hat U_i|\hat U^{i-1}, \hat Y^n).
\end{align*}
\end{Proposition}
\medskip

\begin{IEEEproof}[Proof of Lemma~\ref{lemmaPolar}]
Using Propositions~\ref{propositionPolar1} and~\ref{propositionPolar2} and following similar procedures in the proof of~\cite[Th.~1]{HondaYamamoto13}, we obtain Lemma~\ref{lemmaPolar}.
\end{IEEEproof}

\section{Proof of Lemma~\ref{lemmaVD}} \label{appendixB}
Fix a $p_X$ and a $k\in\mathbb{N}$, and let $n=2^k$. Let $r_{U^n, X^n, Y^n}$ be as defined in~\eqref{defDistR}, which is the distribution induced by the randomized $(n, p_X,  \mathcal{I}, \Lambda_{\mathcal{I}^c})$-polar code where $\Lambda_{\mathcal{I}^c}\sim r_{\Lambda_{\mathcal{I}^c}}$. Let $p_{U^n, X^n, Y^n}$ be the distribution as defined in~\eqref{defP}. In this proof some subscripts of distributions are omitted for simplicity. In order to prove~\eqref{lemmaVDstatement}, we consider the following chain of inequalities:
\begin{align}
 2\|p_{U^n, X^n, Y^n} - r_{U^n, X^n, Y^n}\| 
& = \sum_{u^n\in \mathcal{U}^n, x^n\in \mathcal{X}^n, y^n\in \mathcal{Y}^n}|p(u^n, x^n, y^n)-r(u^n, x^n, y^n)|\notag\\
&  =\hspace{-0.4 in}\sum_{\qquad \quad  u^n\in \mathcal{U}^n, x^n\in \mathcal{X}^n, y^n\in \mathcal{Y}^n}\hspace{-0.45 in}|p(u^n, y^n)p(x^n|u^n, y^n)-r(u^n, y^n)r(x^n|u^n, y^n)|\notag\\
&  \stackrel{\text{(a)}}{=}\hspace{-0.4 in}\sum_{\qquad \quad u^n\in \mathcal{U}^n, x^n\in \mathcal{X}^n, y^n\in \mathcal{Y}^n}\hspace{-0.4 in}|p(u^n, y^n)p(x^n|u^n)-r(u^n, y^n)p(x^n|u^n)|\notag\\
& \stackrel{\text{(b)}}{=}\sum_{u^n\in \mathcal{U}^n, y^n\in \mathcal{Y}^n}|p(u^n, y^n)-r(u^n, y^n)|\notag\\
&\stackrel{\text{(c)}}{=} \sum_{u^n\in \mathcal{U}^n, y^n\in \mathcal{Y}^n}|p(u^n)-r(u^n)|p(y^n|u^n)\notag\\
& = \sum_{u^n\in \mathcal{U}^n}|p(u^n)-r(u^n)|\notag\\
& = \sum_{u^n\in \mathcal{U}^n} \Bigg|\sum_{i=1}^n \left(p(u_i|u^{i-1})-r(u_i|u^{i-1})\right) \left(\prod_{\ell=1}^{i-1}p(u_\ell|u^{\ell-1})\right)\left(\prod_{\ell=i+1}^n r(u_\ell|u^{\ell-1})\right)\Bigg|\notag\\
&\le  \sum_{i=1}^n \sum_{u^i\in \mathcal{U}^i} \left|p(u_i|u^{i-1})-r(u_i|u^{i-1})\right|\prod_{\ell=1}^{i-1}p(u_\ell|u^{\ell-1}) \label{lemmaVDproofEq1}
\end{align}
where
\begin{enumerate}
\item[(a)] follows from \eqref{defpUgivenXinPolar} and the fact by~\eqref{defGn} that $G_n$ is invertible.
\item[(b)] follows from the fact by \eqref{defpUgivenX} that for each $u^n\in \mathcal{U}^n$, there exists an $x^n\in\mathcal{X}^n$ such that $p(x^n|u^n)=1$.
    \item[(c)] follows from the fact by~\eqref{defpUgivenXinPolar} that given $u^n$,
    \begin{equation*}
    p_{Y^n|U^n=u^n}(y^n)=r_{Y^n|U^n=u^n}(y^n) = \prod_{i=1}^nq_{Y|X}(y_i|\tilde x_i)
    \end{equation*}
    where
    \begin{equation*}
    [\tilde x_1\ \tilde x_2\ \ldots \tilde x_n] = [u_1\ u_2\ \ldots u_n]G_n^{-1}.
    \end{equation*}
\end{enumerate}
Using Definition~\ref{defPolarCode} and recalling that $r_{U^n, X^n, Y^n|\Lambda_{\mathcal{I}^c}=\lambda_{\mathcal{I}^c}}$ is the distribution induced by the $(n, p_X, \mathcal{I}, \lambda_{\mathcal{I}^c})$-polar code, we have
\begin{equation}
r(u^i|\lambda_{\mathcal{I}^c})= r(u^i|\lambda_{\mathcal{I}^c\cap\{1, 2, \ldots, i\}}) \label{lemmaVDproofEq2}
\end{equation}
for each $i\in \{1, 2, \ldots, n\}$
and
\begin{equation}
r(u_i|u^{i-1},\lambda_{\mathcal{I}^c})= r(u_i|u^{i-1},\lambda_i) \label{lemmaVDproofEq3}
\end{equation}
for each $i\in \mathcal{I}^c$. Following~\eqref{lemmaVDproofEq1}, we consider for each $i\in\mathcal{I}^c$ and each $u^i\in\{0,1\}^i$
    \begin{align}
   r(u_i|u^{i-1}) 
  & \stackrel{\eqref{defDistR}}{=} \frac{ \sum\limits_{\lambda_{\mathcal{I}^c}\in \Gamma(\mathcal{I}^c)}
r(\lambda_{\mathcal{I}^c})r(u^i|\lambda_{\mathcal{I}^c})}{\sum\limits_{\lambda_{\mathcal{I}^c}\in \Gamma(\mathcal{I}^c)}
r(\lambda_{\mathcal{I}^c})r(u^{i-1}|\lambda_{\mathcal{I}^c})}\notag\\
&\stackrel{\text{(a)}}{=}\frac{ \sum\limits_{\lambda_{\mathcal{I}^c}\in \Gamma(\mathcal{I}^c)}
r(\lambda_{\mathcal{I}^c})r(u^{i-1}|\lambda_{\mathcal{I}^c\cap\{1,2, \ldots, i-1\}})r(u_i|u^{i-1}, \lambda_i)}{\sum\limits_{\lambda_{\mathcal{I}^c}\in \Gamma(\mathcal{I}^c)}
r(\lambda_{\mathcal{I}^c})r(u^{i-1}|\lambda_{\mathcal{I}^c\cap\{1, 2, \ldots, i-1\}})}\notag\\
&\stackrel{\eqref{defDistRprod}}{=} \frac{\hspace{-0.5 in}\sum\limits_{\qquad\qquad \boldsymbol{\lambda}\in \Gamma(\mathcal{I}^c)\cap\{1, 2, \ldots, i-1\}}\hspace{-0.5 in}
r(\boldsymbol{\lambda})r(u^{i-1}|\boldsymbol{\lambda})\sum\limits_{\lambda_i\in \Gamma(\{i\})}r(\lambda_i) r(u_i|u^{i-1}, \lambda_i)}{\sum\limits_{\boldsymbol{\lambda}\in \Gamma(\mathcal{I}^c)\cap\{1, 2, \ldots, i-1\}}
r(\boldsymbol{\lambda})r(u^{i-1}|\boldsymbol{\lambda})}\notag\\
& \stackrel{\eqref{defLambdaI}}{=} \frac{\hspace{-0.65 in}\sum\limits_{\quad\qquad \qquad\boldsymbol{\lambda}\in \Gamma(\mathcal{I}^c)\cap\{1, 2, \ldots, i-1\}}
\hspace{-0.65 in}r(\boldsymbol{\lambda})r(u^{i-1}|\boldsymbol{\lambda})\sum\limits_{\lambda_i\in \Gamma(\{i\})}r(\lambda_i) \mathbf{1}\{u_i = \lambda_i(u^{i-1})\}}{\sum\limits_{\boldsymbol{\lambda}\in \Gamma(\mathcal{I}^c)\cap\{1, 2, \ldots, i-1\}}
r(\boldsymbol{\lambda})r(u^{i-1}|\boldsymbol{\lambda})}\notag\\*
&\stackrel{\text{(b)}}{=} p_{U_i|U^{i-1}}(u_i|u^{i-1}), \label{lemmaVDproofEq4}
    \end{align}
 where
 \begin{enumerate}
 \item[(a)] follows from \eqref{lemmaVDproofEq2} and~\eqref{lemmaVDproofEq3}.
 \item[(b)] follows from the fact by \eqref{defDistRlambdaI} that for each $u^{i-1}\in\{0,1\}^{i-1}$
 \begin{equation*}
 \Pr_{r_{\Lambda_i}}\left\{u_i=\Lambda_i(u^{i-1})\right\} = p_{U_i|U^{i-1}}(u_i|u^{i-1})
 \end{equation*}
 for each $u_i\in\{0,1\}$.
 \end{enumerate}
 Combining~\eqref{lemmaVDproofEq1} and~\eqref{lemmaVDproofEq4}, we obtain
 \begin{align}
 2\|p_{U^n, X^n, Y^n} - r_{U^n, X^n, Y^n}\|  \le \sum_{i\in \mathcal{I}} \sum_{u^i\in \mathcal{U}^i} \left|p(u_i|u^{i-1})-r(u_i|u^{i-1})\right|p(u^{i-1}). \label{lemmaVDproofEq5}
 \end{align}
For each $i\in \mathcal{I}$, since
\begin{align}
 \sum_{u^i\in \mathcal{U}^i} \left|p(u_i|u^{i-1})-r(u_i|u^{i-1})\right|p(u^{i-1})
 & \stackrel{\eqref{defUniformBits}}{=} \sum_{u^{i-1}\in \mathcal{U}^{i-1}}p(u^{i-1}) \sum_{u_i\in \{0,1\}} \left|p(u_i|u^{i-1})-1/2\right|\notag\\
&\stackrel{\text{(a)}}{\le} \sum_{u^{i-1}\in \mathcal{U}^{i-1}}p(u^{i-1}) \sqrt{2\ln 2(1-H_{p_{U_i|U^{i-1}=u^{i-1}}}(U_i))}\notag\\
&\stackrel{\text{(b)}}{\le} \sqrt{2\ln 2(1-H_{p_{U^i}}(U_i|U^{i-1}))}\notag\\
&\stackrel{\eqref{bhattacharyyaEntropy}}{\le} \sqrt{2\ln 2\left(1-(Z_{p_{U^i}}(U_i|U^{i-1}))^2\right)}\notag\\
&\stackrel{\eqref{defInformationBitSet}}{\le}  \sqrt{2\ln 2\left(1-\left(1-n^{-4}\right)^2\right)}\notag\\
&\le \frac{2\sqrt{\ln 2}}{n^2} \notag
\end{align}
where (a) follows from Pinsker's inequality and (b) follows from Jensen's inequality, it follows from~\eqref{lemmaVDproofEq5} that
\begin{align*}
\|p_{U^n, X^n, Y^n} - r_{U^n, X^n, Y^n}\| \le \frac{\sqrt{\ln 2}}{n}.
\end{align*}

\begin{thebibliography}{10}
\providecommand{\url}[1]{#1}
\csname url@samestyle\endcsname
\providecommand{\newblock}{\relax}
\providecommand{\bibinfo}[2]{#2}
\providecommand{\BIBentrySTDinterwordspacing}{\spaceskip=0pt\relax}
\providecommand{\BIBentryALTinterwordstretchfactor}{4}
\providecommand{\BIBentryALTinterwordspacing}{\spaceskip=\fontdimen2\font plus
\BIBentryALTinterwordstretchfactor\fontdimen3\font minus
  \fontdimen4\font\relax}
\providecommand{\BIBforeignlanguage}[2]{{%
\expandafter\ifx\csname l@#1\endcsname\relax
\typeout{** WARNING: IEEEtran.bst: No hyphenation pattern has been}%
\typeout{** loaded for the language `#1'. Using the pattern for}%
\typeout{** the default language instead.}%
\else
\language=\csname l@#1\endcsname
\fi
#2}}
\providecommand{\BIBdecl}{\relax}
\BIBdecl

\bibitem{FTY15}
S.~L. Fong, V.~Y.~F. Tan, and J.~Yang, ``Non-asymptotic achievable rates for
  energy-harvesting channels using save-and-transmit,'' \emph{{IEEE} J. Sel.
  Areas Commun.}, vol.~34, no.~12, 2016.

\bibitem{Ozel:2012:AWGN}
O.~Ozel and S.~Ulukus, ``Achieving {AWGN} capacity under stochastic energy
  harvesting,'' \emph{IEEE Transactions on Information Theory}, vol.~58,
  no.~10, pp. 6471--6483, 2012.

\bibitem{ulukus15}
S.~Ulukus, A.~Yener, E.~Erkip, O.~Simeone, M.~Zorzi, P.~Grover, and K.~Huang,
  ``Energy harvesting wireless communications: A review of recent advances,''
  \emph{{IEEE} J. Sel. Areas Commun.}, vol.~33, no.~3, pp. 360--381, 2015.

\bibitem{MHU15}
M.~Mondelli, S.~H. Hassani, and R.~Urbanke, ``Unified scaling of polar codes:
  {Error} exponent, scaling exponent, moderate deviations, and error floors,''
  \emph{to appear in IEEE Trans. Inf. Theory}, 2016, {\tt arXiv:1501.02444
  [cs.IT]}.

\bibitem{HAU14}
S.~H. Hassani, K.~Alishahi, and R.~Urbanke, ``Finite-length scaling for polar
  codes,'' \emph{{IEEE} Trans. Inf. Theory}, vol.~60, no.~10, pp. 5875--5898,
  2014.

\bibitem{GoldinBurshtein14}
D.~Goldin and D.~Burshtein, ``Improved bounds on the finite length scaling of
  polar codes,'' \emph{{IEEE} Trans. Inf. Theory}, vol.~60, no.~11, pp.
  6966--6978, 2014.

\bibitem{GoldinBurshtein15}
------, ``On the finite length scaling of ternary polar codes,'' in \emph{Proc.
  IEEE Intl. Symp. Inf.~Theory}, Hong Kong, Jun. 2015.

\bibitem{STA09}
E.~\c{S}a\c{s}o\u{g}lu, I.~Telatar, and E.~Ar{\i}kan, ``Polarization of
  arbitrary discrete memoryless channels,'' in \emph{Proc.~IEEE Inf.~Theory
  Workshop}, Seoul, Korea, Oct. 2009, pp. 114--118.

\bibitem{SRDR12}
D.~Sutter, J.~M. Renes, F.~Dupuis, and R.~Renner, ``Achieving the capacity of
  any {DMC} using only polar codes,'' in \emph{Proc.~IEEE Inf.~Theory
  Workshop}, Lausanne, Switzerland, Sep. 2012, pp. 114--118.

\bibitem{HondaYamamoto13}
J.~Honda and H.~Yamamoto, ``Polar coding without alphabet extension for
  asymmetric models,'' \emph{{IEEE} Trans. Inf. Theory}, vol.~59, no.~12, pp.
  7829--7838, 2013.

\bibitem{MHU14Allerton}
M.~Mondelli, R.~Urbanke, and S.~H. Hassani, ``{How to Achieve the Capacity of
  Asymmetric Channels},'' in \emph{{Proc.~Allerton Conference on Communication,
  Control and Computing}}, Oct. 2014, pp. 789--796.

\bibitem{AbbeBarron11}
E.~Abbe and A.~Barron, ``Polar coding schemes for the {AWGN} channel,'' in
  \emph{Proc. IEEE Intl. Symp. Inf.~Theory}, St Petersburg, Russia, Jul. 2011,
  pp. 194--198.

\bibitem{elgamal}
A.~{El~Gamal} and Y.-H. Kim, \emph{Network Information Theory}.\hskip 1em plus
  0.5em minus 0.4em\relax Cambridge, U.K.: Cambridge University Press, 2012.

\bibitem{Arikan:10ISIT}
E.~{Ar\i kan}, ``Source polarization,'' in \emph{Proc. IEEE Intl. Symp.
  Inf.~Theory}, Austin, TX, USA, Jun. 2010, pp. 899--903.

\bibitem{PPV10}
Y.~Polyanskiy, H.~V. Poor, and S.~Verd\'{u}, ``Channel coding rate in the
  finite blocklength regime,'' \emph{{IEEE} Trans. Inf. Theory}, vol.~56,
  no.~5, pp. 2307--2359, 2010.

\bibitem{Strassen}
V.~Strassen, ``Asymptotische absch\"{a}tzungen in {Shannons}
  informationstheorie,'' \emph{Trans. Third Prague Conf. Inf. Theory}, pp.
  689--723, 1962, http://www.math.cornell.edu/\texttildelow pmlut/strassen.pdf.

\bibitem{Hayashi09}
M.~Hayashi, ``Information spectrum approach to second-order coding rate in
  channel coding,'' \emph{{IEEE} Trans. Inf. Theory}, vol.~55, no.~11, pp.
  4947--4966, 2009.

\bibitem{arikan09}
E.~{Ar\i kan}, ``Channel polarization: {A} method for constructing
  capacity-achieving codes for symmetric binary-input memoryless channels,''
  \emph{{IEEE} Trans. Inf. Theory}, vol.~55, no.~7, pp. 3051--3073, 2009.

\end{thebibliography}
\end{document}